\documentclass[longauth]{aa}
\usepackage{xspace}
\usepackage{color}
\usepackage{lineno}
\usepackage{graphics}
\usepackage{txfonts}
\usepackage{subcaption}
\usepackage[section]{placeins}
\usepackage{graphicx}
\usepackage[export]{adjustbox}
\usepackage{natbib}
\usepackage{deluxetable}
\usepackage{amsmath, amssymb}
\usepackage{stfloats}
\usepackage{hyperref}
\usepackage{deluxetable}
\usepackage{longtable}
\usepackage[para]{footmisc}

\usepackage{caption}
\begin{document} 

    \title{Giant Outer Transiting Exoplanet Mass (GOT\,’EM) Survey.VII. TOI-6041: a multi-planet system including a warm Neptune exhibiting strong TTVs}
    \author{
N.~Heidari\inst{\ref{inst1}}\thanks{E-mail:heidari@iap.fr}
\and
A.~Alnajjarine\inst{\ref{inst2}}
\and
H.~P.~Osborn\inst{\ref{inst3}}\fnmsep\inst{\ref{inst4}}
\and
D.~Dragomir\inst{\ref{inst6}}
\and
P.~Dalba\inst{\ref{inst5}}
\and
W.~Benz\inst{\ref{inst7}}\fnmsep\inst{\ref{inst3}}
\and
G.~H\'ebrard\inst{\ref{inst1}}\fnmsep\inst{\ref{inst9}}
\and
J.~Laskar\inst{\ref{inst2}}
\and
N.~Billot\inst{\ref{inst8}}
\and
M.~N.~Günther\inst{\ref{inst10}}
\and
T.~G.~Wilson\inst{\ref{inst11}}
\and
Y.~Alibert\inst{\ref{inst3}}\fnmsep\inst{\ref{inst7}}
\and
A.~Bonfanti\inst{\ref{inst12}}
\and
A.~Bieryla\inst{\ref{inst13}}
\and
C.~Broeg\inst{\ref{inst7}}\fnmsep\inst{\ref{inst3}}
\and
A.~C.~M.~Correia\inst{\ref{inst14}}
\and
J.~A.~Egger\inst{\ref{inst7}}
\and
Z.~Essack\inst{\ref{inst6}}
\and
E.~Furlan\inst{\ref{inst15}}
\and
D.~Gandolfi\inst{\ref{inst16}}
\and
N.~Grieves\inst{\ref{inst17}}
\and
S.~Howell\inst{\ref{inst18}}
\and
D.~LaCourse\inst{\ref{inst19}}
\and
C.~Pezzotti\inst{\ref{inst20}}
\and
T.~Pritchard \inst{\ref{inst23}}
\and
S.~G.~Sousa\inst{\ref{inst21}}
\and
S.~Ulmer-Moll\inst{\ref{inst22}}\fnmsep\inst{\ref{inst20}}
\and
S.~Villanueva\inst{\ref{inst23}}
\and
R.~Alonso\inst{\ref{inst24}}\fnmsep\inst{\ref{inst25}}
\and
J.~Asquier\inst{\ref{inst10}}
\and
T.~Bárczy\inst{\ref{inst26}}
\and
D.~Barrado\inst{\ref{inst27}}
\and
S.~C.~C.~Barros\inst{\ref{inst21}}\fnmsep\inst{\ref{inst28}}
\and
W.~Baumjohann\inst{\ref{inst12}}
\and
L.~Borsato\inst{\ref{inst29}}
\and
A.~Brandeker\inst{\ref{inst30}}
\and
M.~Buder\inst{\ref{inst31}}
\and
A.~Collier Cameron\inst{\ref{inst32}}
\and
S.~Csizmadia\inst{\ref{inst31}}
\and
P.~E.~Cubillos\inst{\ref{inst12}}\fnmsep\inst{\ref{inst33}}
\and
M.~B.~Davies\inst{\ref{inst34}}
\and
M.~Deleuil\inst{\ref{inst35}}
\and
X.~Delfosse\inst{\ref{inst36}}
\and
A.~Deline\inst{\ref{inst8}}
\and
O.~D.~S.~Demangeon\inst{\ref{inst21}}\fnmsep\inst{\ref{inst28}}
\and
B.~Demory\inst{\ref{inst3}}\fnmsep\inst{\ref{inst37}}\fnmsep\inst{\ref{inst7}}
\and
A.~Derekas\inst{\ref{inst38}}
\and
B.~Edwards\inst{\ref{inst39}}
\and
D.~Ehrenreich\inst{\ref{inst8}}\fnmsep\inst{\ref{inst40}}
\and
A.~Erikson\inst{\ref{inst31}}
\and
A.~Fortier\inst{\ref{inst7}}\fnmsep\inst{\ref{inst3}}
\and
L.~Fossati\inst{\ref{inst12}}
\and
M.~Fridlund\inst{\ref{inst41}}\fnmsep\inst{\ref{inst42}}
\and
K.~Gazeas\inst{\ref{inst43}}
\and
M.~Gillon\inst{\ref{inst44}}
\and
M.~Güdel\inst{\ref{inst45}}
\and
J.~Hasiba\inst{\ref{inst12}}
\and
A.~Heitzmann\inst{\ref{inst8}}
\and
C.~Helling\inst{\ref{inst12}}\fnmsep\inst{\ref{inst46}}
\and
J.~M.~Jenkins\inst{\ref{inst18}}
\and
T.~Keller\inst{\ref{inst7}}\fnmsep\inst{\ref{inst3}}
\and
K.~G.~Stassun\inst{\ref{inst47}}
\and
L.~L.~Kiss\inst{\ref{inst48}}\fnmsep\inst{\ref{inst49}}
\and
J.~Korth\inst{\ref{inst8}}
\and
K.~W.~F.~Lam\inst{\ref{inst31}}
\and
D.~W.~Latham\inst{\ref{inst13}}
\and
A.~Lecavelier des Etangs\inst{\ref{inst50}}
\and
A.~Leleu\inst{\ref{inst8}}\fnmsep\inst{\ref{inst7}}
\and
M.~Lendl\inst{\ref{inst8}}
\and
P.~F.~L.~Maxted\inst{\ref{inst51}}
\and
S. ~McDermott\inst{\ref{inst65}}
\and
B.~Merín\inst{\ref{inst52}}
\and
C.~Mordasini\inst{\ref{inst7}}\fnmsep\inst{\ref{inst3}}
\and
V.~Nascimbeni\inst{\ref{inst29}}
\and
G.~ Nowak\inst{\ref{inst53}}
\and
G.~Olofsson\inst{\ref{inst30}}
\and
I.~Pagano\inst{\ref{inst54}}
\and
E.~Pallé\inst{\ref{inst24}}\fnmsep\inst{\ref{inst25}}
\and
G.~Piotto\inst{\ref{inst29}}\fnmsep\inst{\ref{inst55}}
\and
D.~Pollacco\inst{\ref{inst11}}
\and
D.~Queloz\inst{\ref{inst4}}\fnmsep\inst{\ref{inst56}}
\and
R.~Ragazzoni\inst{\ref{inst29}}\fnmsep\inst{\ref{inst55}}
\and
H.~Rauer\inst{\ref{inst57}}\fnmsep\inst{\ref{inst58}}
\and
I.~Ribas\inst{\ref{inst59}}\fnmsep\inst{\ref{inst60}}
\and
G.~Ricker\inst{\ref{inst61}}
\and
N.~C.~Santos\inst{\ref{inst21}}\fnmsep\inst{\ref{inst28}}
\and
G.~Scandariato\inst{\ref{inst54}}
\and
S.~Seager\inst{\ref{inst61}}
\and
D.~Ségransan\inst{\ref{inst8}}
\and
A.~E.~Simon\inst{\ref{inst7}}\fnmsep\inst{\ref{inst3}}
\and
A.~M.~S.~Smith\inst{\ref{inst31}}
\and
M.~Stalport\inst{\ref{inst20}}\fnmsep\inst{\ref{inst44}}
\and
S.~Striegel\inst{\ref{inst66}}
\and
S.~Sulis\inst{\ref{inst35}}
\and
G.~M.~Szabó\inst{\ref{inst38}}\fnmsep\inst{\ref{inst62}}
\and
S.~Udry\inst{\ref{inst8}}
\and
V.~Van Grootel\inst{\ref{inst20}}
\and
R.~Vanderspek\inst{\ref{inst61}}
\and
J.~Venturini\inst{\ref{inst8}}
\and
E.~Villaver\inst{\ref{inst24}}\fnmsep\inst{\ref{inst25}}
\and
V.~Viotto\inst{\ref{inst29}}
\and
N.~A.~Walton\inst{\ref{inst63}}
\and
J.~N.~Winn\inst{\ref{inst64}}
\and
S.~Wolf\inst{\ref{inst29}}
}

    \institute{
Institut d’astrophysique de Paris, CNRS \& Sorbonne Université, UMR 7095, 98bis
boulevard Arago, 75014 Paris, France\label{inst1} 
\and
LTE, UMR8255 CNRS, Observatoire de Paris, PSL Univ., Sorbonne Univ., 77 av. Denfert-Rochereau, 75014 Paris, France\label{inst2} 
\and
Center for Space and Habitability, University of Bern, Gesellschaftsstrasse 6, 3012 Bern, Switzerland\label{inst3} 
\and
ETH Zurich, Department of Physics, Wolfgang-Pauli-Strasse 2, CH-8093 Zurich, Switzerland\label{inst4} 
\and
Department of Physics and Astronomy, University of New Mexico, 210 Yale Blvd NE, Albuquerque, NM 87106, USA\label{inst6}
\and
Department of Astronomy and Astrophysics, University of California, Santa Cruz, CA 95064, USA\label{inst5} 
\and
Space Research and Planetary Sciences, Physics Institute, University of Bern, Gesellschaftsstrasse 6, 3012 Bern, Switzerland\label{inst7} 
\and
Observatoire astronomique de l'Université de Genève, Chemin Pegasi 51, 1290 Versoix, Switzerland\label{inst8} 
\and
Observatoire de Haute-Provence, CNRS, Universit\'e d'Aix-Marseille, 04870 Saint-Michel-l'Observatoire, France\label{inst9} 
\and
European Space Agency (ESA), European Space Research and Technology Centre (ESTEC), Keplerlaan 1, 2201 AZ Noordwijk, The Netherlands\label{inst10} 
\and
Department of Physics, University of Warwick, Gibbet Hill Road, Coventry CV4 7AL, United Kingdom\label{inst11} 
\and
Space Research Institute, Austrian Academy of Sciences, Schmiedlstrasse 6, A-8042 Graz, Austria\label{inst12} 
\and
Center for Astrophysics \textbar Harvard \& Smithsonian, 60 Garden Street, Cambridge, MA 02138, USA\label{inst13} 
\and
CFisUC, Departamento de Física, Universidade de Coimbra, 3004-516 Coimbra, Portugal\label{inst14} 
\and
Caltech/IPAC-NASA Exoplanet Science Institute, 770 S. Wilson Avenue, Pasadena, CA 91106, USA\label{inst15} 
\and
Dipartimento di Fisica, Università degli Studi di Torino, via Pietro Giuria 1, I-10125, Torino, Italy\label{inst16} 
\and
Observatoire de Genève, Université de Genève, 51 Chemin Pegasi, 1290 Versoix, Switzerland\label{inst17} 
\and
NASA Ames Research Center, Moffett Field, CA 94035, USA\label{inst18} 
\and
Amateur Astronomer, 7507 52nd Place NE, Marysville, WA 98270, USA\label{inst19} 
\and
Space sciences, Technologies and Astrophysics Research (STAR) Institute, Université de Liège, Allée du 6 Août 19C, 4000 Liège, Belgium\label{inst20} 
\and
Instituto de Astrofisica e Ciencias do Espaco, Universidade do Porto, CAUP, Rua das Estrelas, 4150-762 Porto, Portugal\label{inst21} 
\and
Leiden Observatory, University of Leiden, Einsteinweg 55, 2333 CA Leiden, The Netherlands\label{inst22} 
\and
NASA Goddard Space Flight Center, 8800 Greenbelt Rd, Greenbelt, MD 20771, USA\label{inst23} 
\and
Instituto de Astrofísica de Canarias, Vía Láctea s/n, 38200 La Laguna, Tenerife, Spain\label{inst24} 
\and
Departamento de Astrofísica, Universidad de La Laguna, Astrofísico Francisco Sanchez s/n, 38206 La Laguna, Tenerife, Spain\label{inst25} 
\and
Admatis, 5. Kandó Kálmán Street, 3534 Miskolc, Hungary\label{inst26} 
\and
Depto. de Astrofísica, Centro de Astrobiología (CSIC-INTA), ESAC campus, 28692 Villanueva de la Cañada (Madrid), Spain\label{inst27} 
\and
Departamento de Fisica e Astronomia, Faculdade de Ciencias, Universidade do Porto, Rua do Campo Alegre, 4169-007 Porto, Portugal\label{inst28} 
\and
INAF, Osservatorio Astronomico di Padova, Vicolo dell'Osservatorio 5, 35122 Padova, Italy\label{inst29} 
\and
Department of Astronomy, Stockholm University, AlbaNova University Center, 10691 Stockholm, Sweden\label{inst30} 
\and
Institute of Space Research, German Aerospace Center (DLR), Rutherfordstrasse 2, 12489 Berlin, Germany\label{inst31} 
\and
Centre for Exoplanet Science, SUPA School of Physics and Astronomy, University of St Andrews, North Haugh, St Andrews KY16 9SS, UK\label{inst32} 
\and
INAF, Osservatorio Astrofisico di Torino, Via Osservatorio, 20, I-10025 Pino Torinese To, Italy\label{inst33} 
\and
Centre for Mathematical Sciences, Lund University, Box 118, 221 00 Lund, Sweden\label{inst34} 
\and
Aix Marseille Univ, CNRS, CNES, LAM, 38 rue Frédéric Joliot-Curie, 13388 Marseille, France\label{inst35} 
\and
Univ. Grenoble Alpes, CNRS, IPAG, 38000 Grenoble, France\label{inst36} 
\and
ARTORG Center for Biomedical Engineering Research, University of Bern, Bern, Switzerland\label{inst37} 
\and
ELTE Gothard Astrophysical Observatory, 9700 Szombathely, Szent Imre h. u. 112, Hungary\label{inst38} 
\and
SRON Netherlands Institute for Space Research, Niels Bohrweg 4, 2333 CA Leiden, Netherlands\label{inst39} 
\and
Centre Vie dans l’Univers, Faculté des sciences, Université de Genève, Quai Ernest-Ansermet 30, 1211 Genève 4, Switzerland\label{inst40} 
\and
Leiden Observatory, University of Leiden, PO Box 9513, 2300 RA Leiden, The Netherlands\label{inst41} 
\and
Department of Space, Earth and Environment, Chalmers University of Technology, Onsala Space Observatory, 439 92 Onsala, Sweden\label{inst42} 
\and
National and Kapodistrian University of Athens, Department of Physics, University Campus, Zografos GR-157 84, Athens, Greece\label{inst43} 
\and
Astrobiology Research Unit, Université de Liège, Allée du 6 Août 19C, B-4000 Liège, Belgium\label{inst44} 
\and
Department of Astrophysics, University of Vienna, Türkenschanzstrasse 17, 1180 Vienna, Austria\label{inst45} 
\and
Institute for Theoretical Physics and Computational Physics, Graz University of Technology, Petersgasse 16, 8010 Graz, Austria\label{inst46} 
\and
Vanderbilt University, Department of Physics \& Astronomy, 6301 Stevenson Center Ln., Nashville, TN 37235, USA\label{inst47} 
\and
Konkoly Observatory, Research Centre for Astronomy and Earth Sciences, 1121 Budapest, Konkoly Thege Miklós út 15-17, Hungary\label{inst48} 
\and
ELTE E\"otv\"os Lor\'and University, Institute of Physics, P\'azm\'any P\'eter s\'et\'any 1/A, 1117 Budapest, Hungary\label{inst49} 
\and
Institut d'astrophysique de Paris, UMR7095 CNRS, Université Pierre \& Marie Curie, 98bis blvd. Arago, 75014 Paris, France\label{inst50} 
\and
Astrophysics Group, Lennard Jones Building, Keele University, Staffordshire, ST5 5BG, United Kingdom\label{inst51} 
\and
European Space Agency, ESA - European Space Astronomy Centre, Camino Bajo del Castillo s/n, 28692 Villanueva de la Cañada, Madrid, Spain\label{inst52} 
\and
Institute of Astronomy, Faculty of Physics, Astronomy and 
Informatics, Nicolaus Copernicus University, Grudzi\c{a}dzka 5, 87-100 
Toru\'n, Poland\label{inst53} 
\and
INAF, Osservatorio Astrofisico di Catania, Via S. Sofia 78, 95123 Catania, Italy\label{inst54} 
\and
Dipartimento di Fisica e Astronomia "Galileo Galilei", Università degli Studi di Padova, Vicolo dell'Osservatorio 3, 35122 Padova, Italy\label{inst55} 
\and
Cavendish Laboratory, JJ Thomson Avenue, Cambridge CB3 0HE, UK\label{inst56} 
\and
German Aerospace Center (DLR), Markgrafenstrasse 37, 10117 Berlin, Germany\label{inst57} 
\and
Institut fuer Geologische Wissenschaften, Freie Universitaet Berlin, Malteserstrasse 74-100,12249 Berlin, Germany\label{inst58} 
\and
Institut de Ciencies de l'Espai (ICE, CSIC), Campus UAB, Can Magrans s/n, 08193 Bellaterra, Spain\label{inst59} 
\and
Institut d'Estudis Espacials de Catalunya (IEEC), 08860 Castelldefels (Barcelona), Spain\label{inst60} 
\and
Department of Physics and Kavli Institute for Astrophysics and Space Research, Massachusetts Institute of Technology, Cambridge, MA 02139, USA\label{inst61} 
\and
HUN-REN-ELTE Exoplanet Research Group, Szent Imre h. u. 112., Szombathely, H-9700, Hungary\label{inst62} 
\and
Institute of Astronomy, University of Cambridge, Madingley Road, Cambridge, CB3 0HA, United Kingdom\label{inst63} 
\and
Department of Astrophysical Sciences, Princeton University, 4 Ivy Lane, Princeton, NJ 08544, USA\label{inst64}
\and
Proto-Logic LLC, 1718 Euclid Street NW, Washington, DC 20009, USA\label{inst65}
\and
SETI Institute, Mountain View, CA 94043 USA/NASA Ames Research Center, Moffett Field, CA 94035 USA\label{inst66}
}

\date{Received XX, 2025; accepted XX, 2025}
  \abstract
{
We present the characterization of the TOI-6041 system, a bright ($V = 9.84 \pm 0.03$) G7-type star hosting at least two planets. The inner planet, TOI-6041\,b, is a warm Neptune with a radius of $4.55^{+0.18}_{-0.17}\,R_\oplus$, initially identified as a single-transit event in \textit{TESS} photometry. Subsequent observations with \textit{TESS} and \textit{CHEOPS} revealed additional transits, enabling the determination of its $26.04945^{+0.00033}_{-0.00034}$ d orbital period and the detection of significant transit timing variations (TTVs), exhibiting a peak-to-peak amplitude of about 1 hour. Radial velocity (RV) measurements obtained with the APF spectrographs allow us to place a $3\sigma$ upper mass limit of $28.9\,M_\oplus$ on TOI-6041\,b. In addition, the RV data reveal a second companion, TOI-6041\,c, on an 88 d orbit, with a minimum mass of $0.25\,M_{\mathrm{Jup}}$. A preliminary TTV analysis suggests that the observed variations could be caused by gravitational perturbations from planet\,c; however, reproducing the observed amplitudes requires a relatively high eccentricity of about 0.3 for planet\,c. Our dynamical stability analysis indicates that such a configuration is dynamically viable and places a 1$\sigma$ upper limit on the mass of TOI-6041\,c at $0.8\,M_{\mathrm{Jup}}$. An alternative is the presence of a third, low-mass planet located between planets\,b and\,c, or on an inner orbit relative to planet b—particularly near a mean-motion resonance with planet\,b—which could account for the observed variations. These findings remain tentative, and further RV and photometric observations are essential to better constrain the mass of planet b and to refine the TTV modeling, thereby improving our understanding of the system's dynamical architecture.
}

   \keywords{planets and satellites: detection – techniques: photometric, radial velocities – stars: TOI-6041}
\titlerunning{Long-period planet detection and characterization}
\authorrunning{N. Heidari et al }
\maketitle

\section{Introduction}

Gravitational interactions between planets can induce deviations from strictly periodic transit times—known as transit timing variations \citep[TTVs;][]{agol2005detecting,holman2005use}. These variations provide a powerful dynamical tool for estimating planetary masses and probing orbital architectures. For instance, TTVs were first used together with radial velocity (RV) measurements to characterize the Kepler-9 system \citep{2010Sci33051H}, and later enabled the inference of planetary masses from TTVs alone in the Kepler-11 system \citep{2011Natur.470...53L}. In addition, TTVs can be used to confirm planetary candidates or to infer the presence of non-transiting companions, with notable examples including KOI-142 and K2-146 \citep{nesvorny2013koi,hamann2019k2}.

The NASA \textit{Kepler} mission, followed by its \textit{K2} extension, provided the first large statistical sample of exoplanetary systems with detectable TTVs. This enabled numerous dynamical mass measurements and constraints on mutual inclinations \citep{holczer2016transit,hadden2017kepler}. Whereas a few TTV-systems were discovered from the ground \citep[e.g. WASP-148,][]{hebrard2020discovery}, more recently, the NASA \textit{TESS} mission \citep{ricker2015transiting} has expanded the catalog of TTV-bearing systems \citep[e.g.,][]{hobson2023toi,heidari2025characterization}. However, its shorter observing windows—typically around 27 d— limit its sensitivity to TTVs, particularly for planets with longer orbital periods ($P \gtrsim 20$ d). In a given \textit{TESS} sector, such planets often appear as single-transit events. The extended mission revisits the same sectors after about two years and occasionally captures a second transit—producing a so-called “duotransit.” This sparse coverage is usually insufficient for unambiguous period determination or robust TTV detection. Fully characterizing these systems typically requires extensive follow-up, including additional photometric monitoring and RV observations.

In this context, we present the detection and characterization of the TOI-6041 system, a multi-planetary system hosting at least two planets. The inner planet was initially identified as a single-transit event in the \textit{TESS} data. Subsequent \textit{TESS} observations, together with our follow-up campaign—including space-based photometry, high-resolution imaging, and RV measurements—have enabled the characterization of both planets in the system and revealed strong TTVs for the inner planet.

The structure of this paper is as follows. In Sect.~\ref{observation}, we present the observational datasets, including \textit{TESS} and \textit{CHEOPS} photometry, RV follow-up, and high-resolution imaging. Section~\ref{subsec:spectralanalysis} describes the stellar characterization. In Sect.~\ref{detection}, we detail the photometric and spectroscopic analysis. The dynamical modeling and TTV analysis are discussed in Sect.~\ref{dynamics}. Finally, in Sect.~\ref{discussion}, we highlight prospects for further follow-up and summarize our conclusions.

\section{Observations}
\label{observation}
\subsection{TESS photometry}
\label{tess_observation}
\textit{TESS} observed TOI-6041 in three sectors: Sector 18 (November 2019), Sector 58 (November 2022), and Sector 85 (November 2024). A single-transit event was identified in Sector 18 and vetted by the TESS Single Transit Planetary Candidate (TSTPC) group \citep{harris2023separated,burt2021toi} and the Planet Hunters TESS (PHT) citizen scientists \citep{eisner2021planet}. After the second transit was detected in Sector 58, the TESS Science Processing Operations Center (SPOC) assigned the system a TESS Object of Interest (TOI) number \citep{guerrero2021tess}, as the planetary candidate signal successfully passed all diagnostic tests outlined in the Data Validation report \citep{Twicken:DVdiagnostics2018,2019PASP131b4506L}. These tests included the odd-even transit depth test, the ghost diagnostic test, and the difference-image centroiding test. Based on centroiding analyses from Sectors 18 through 85, the transit signal is located within $1.49 \pm 2.53$ arcseconds of the host star TOI-6041.

The two transits observed in Sectors 18 and 58 are separated by 1094.06 d. Multiple possible periods are compatible with the detection of the two transits, from the $P_{\rm min}=20.260$\,d (defined from the lack of transits seen in \textit{TESS} photometry) to $P_{\rm max}=1094.06$\,d (the time interval between the transits). A third transit was later detected in Sector 85, 729 d after the second. It enabled a large number of aliases to be ruled out; however, because it occurred $P_{\rm max}/3 = 364.7$\,d after the previous transit, it was insufficient to uniquely determine the period due to the large time gaps and the limited number of observed events in the \textit{TESS} data alone. Combining the three \textit{TESS} detections with the non-detections in the light curves therefore leaves a minimum allowed period of 20.260 d and a maximum of 364.7 d.

Nevertheless, such candidates offer a simple avenue for photometric period assessment by targeting the times of transit according to each of the possible periods. This has been performed extensively on long-period targets from the ground \citep[e.g.][, etc.]{schanche2022toi,Ulmer-Moll2022} and from space \citep[e.g.][etc.]{osborn2022uncovering,luque2023resonant}.

The data analyzed in this paper include 2-minute cadence observations processed by the Presearch Data Conditioning–Simple Aperture Photometry (PDC-SAP) pipeline \citep{Stumpe2012,stumpe2014multiscale,2012PASP..124.1000S}, provided by the Science Processing Operations Center at NASA Ames Research Center \citep{jenkins2016tess}. The \textit{TESS} light curves were retrieved from the Mikulski Archive for Space Telescopes (MAST\footnote{\url{https://mast.stsci.edu/}}) using the \texttt{lightkurve} package \citep{collaboration2018lightkurve}.

We used \texttt{tpfplotter} \citep{2020AA...635A.128A} to visualize the target pixel files and assess potential light curve contamination, incorporating sources from \textit{Gaia} Data Release 3 \citep[DR3;][]{Gaia2016,Gaia2021}. Our analysis considered sources within the photometric aperture down to a contrast of 6 magnitudes. Beyond this threshold, contamination is negligible, contributing less than 0.4\%. As shown in Fig.~\ref{tpfplotter}, no stars within the \textit{TESS} aperture exceed this 6-magnitude contrast limit in Sector 18. The same applies to Sectors 58 and 85. We therefore did not include dilution from nearby stars in our joint modeling.

Furthermore, using Equation 4 from \citet{2019ApJ...881L..19V}, we estimate that a neighboring star would need to be within a contrast of $\Delta$mag < 2.3 (at 3$\sigma$) to be capable of producing the observed transit depth. No such bright stars are present in the \textit{TESS} aperture, ruling out the possibility that the transit signal is due to a background eclipsing binary. This conclusion is further supported by our high-resolution imaging, presented in Sect.\ref{high_reso_image}.


\subsection{CHEOPS photometry}
\label{sec:CHEOPS_photometry}
As discussed in Sect. \ref{tess_observation}, the orbital period of TOI-6041\,b remained uncertain based on the \textit{TESS} data alone.
Initial modelling of this \textit{TESS} data using the \texttt{MonoTools} package\footnote{\url{https://github.com/hposborn/MonoTools}} \citep{Osborn2022b} revealed 54 aliases.
More usefully, the fit allows the marginal period probability across the aliases thanks to a combination of the transit model likelihood, a geometric transit probability prior, a window function prior \citep[see ][]{kipping2018orbital}, and an eccentricity prior (in this case using a Beta distribution derived from populations of planets with $2<R_P<6\,R_\oplus$ from Bern model simulations, \citep{Emsenhuber2021}).
The result is shown in Fig. \ref{fig:monotools}.
In order to resolve the true period, we turned to observations with \textit{CHEOPS}.

\textit{CHEOPS} \citep{benz2021cheops,fortier2024cheops} is a 30~cm aperture ESA telescope dedicated to the study of
transiting exoplanets. It performs high-precision optical photometry in a 330--1100 nm passband, from a sun-synchronous \& nadir-locked low-earth orbit. One science case within the \textit{CHEOPS} GTO has been to determine the orbital periods of planetary candidates found to have ambiguous periods in \textit{TESS} photometry \citep[see e.g., ][]{osborn2022uncovering,Tuson2023,psaridi2024discovery}.

TOI-6041 was observed on nine occasions by \textit{CHEOPS} in 2023 \& 2024 (see Table \ref{tab:cheops_dat}).
Due to the deep depth of TOI-6041 b and large expected SNR in \textit{CHEOPS}, we designed an observing strategy in order to flexibly observe partial transits.
This allowed for short (3-orbits, $\sim5$\,hr) observations to be more easily
scheduled between higher-priority observations. An initial transit was found at $\sim$2460254.8 BJD, with the full transit event being observed. 
This transit occurred exactly $P_{\rm max}/2$ after the second \textit{TESS} transit, allowing us to exclude half the aliases. A further transit ingress was clearly seen at $\sim$2460306.9 BJD, which corresponded to two potential periods - either 52.1 \& 26.05\,d. Finally, in order to determine the period, we observed a time corresponding to only the 26.05\,d alias at $\sim$2460645.5 BJD, finding another clear ingress feature. Together, these three transits determined a period of $26.049\pm0.001$\,d for
TOI-6041 b.

Due to \textit{CHEOPS}' field rotation once per 99\.min orbit, substantial detrending is necessary to achieve photometry close to the expected photon noise. We initially processed the subarrays using the \texttt{PIPE} package \footnote{\url{https://github.com/alphapsa/PIPE}}, which models the point spread function (PSF) on the images \citep[see e.g.][]{brandeker2022cheops,morris2021cheops}.
We then used the \texttt{chexoplanet} package
\footnote{\url{https://github.com/hposborn/chexoplanet}} in order to further detrend the \textit{CHEOPS} photometry. 
This first assesses which house-keeping data such as background, centroids, on-board temperature correlate with flux, looking at both linear \& quadratic trends. It then builds a model which decorrelates the flux using these parameters, as well as a spline with $\sim9\,$deg spacing to model rapid changes in flux as a function of spacecraft roll-angle, and a simple transit model \citep[see e.g.][]{egger2024unveiling}.
The result is photometry with a precision of $\sim330$\,ppm$/$hr.

\begin{table*} 
\centering
\resizebox{1.8\columnwidth}{!}{%
\begin{tabular}{lcccccc} 
\hline 
BJD start & Dur [orb] & Filekey & Eff. [\%] & RMS [ppm] & Aliases [d] \\ 
\hline 
$ 2460220.4030 $ & $ 3.18 $ & TG004901 & 59 & $ 339 $ 
& P=20.6 \\  
$ 2460226.5118 $ & $ 2.59 $ & TG005001 & 59 & $ 304 $ 
& P=[84.2,42.1,28.0,21.0] \\ 
$ 2460236.9567 $ & $ 3.32 $ & TG006201 & 59 & $ 328 $ 
& 26.7 \\ 
$ 2460254.5351 $ & $ 2.9 $ & TG004801 & 60 & $ 326 $ 
& 18 aliases\\ 
$ 2460254.7455 $ & $ 2.9 $ & TG005101 &55& $ 475 $ 
& \textbf{18 aliases}\\ 
$ 2460261.8290 $ & $ 2.9 $ & TG005201 &56& $ 295 $ 
& P=20.6 \\ 
$ 2460262.5632 $ & $ 2.56 $ & TG005501 &60& $ 359 $ 
& P=23.3 \\ 
$ 2460306.6604 $ & $ 3.28 $ & TG006101 & 56& $ 333 $ 
& \textbf{P=[26.0,52.1]}\\ 
$ 2460645.3298 $ & $ 3.07 $ & TG030701 &53& $ 425 $ 
& \textbf{P=26.0}\\
\hline 
\end{tabular}%
}
\caption{Summary of \textit{CHEOPS} observations. Each duration is given in units of \textit{CHEOPS} orbits (1 orbit = 98.99 minutes). Efficiency is averaged between start- and end-of visit and the RMS is reported in ppm per hour. Filekeys are all preceeded by \texttt{PR140079\_} and succeeded by \texttt{\_V0300}. The "Aliases" column lists orbital period aliases that are consistent with a transit occurring at the specified BJD. Rows in bold indicate observations where the transit of TOI-6041\,b was successfully detected.} 
\label{tab:cheops_dat}
\end{table*}

\subsection{APF spectroscopy}

We collected 102 RV measurements of TOI-6041 using the Levy High-Resolution Spectrograph ($R \approx 114{,}000$; \citealt{burt2014achieving}) on the 2.4\,m Automated Planet Finder (APF) telescope at Lick Observatory \citep{vogt2014apf,radovan2014automated,radovan2010radial}. Observations were conducted between January 30, 2020, and June 27, 2023. The RVs have a median uncertainty of 2.6\,m\,s$^{-1}$ and a mean dispersion of 13.2\,m\,s$^{-1}$. For further details on the instrumental setup and data reduction, see \citet{fulton2015three}.

Four measurements were excluded due to their high error bars. The complete list of corrected RVs is provided in Table~\ref{tab:rvs APF}.

\subsection{SOPHIE spectroscopy}

We observed TOI-6041 using the SOPHIE high-resolution spectrograph, mounted on the 193 cm telescope at the Observatoire de Haute-Provence \citep[OHP;][]{perruchot2008sophie, bouchy2013sophie+}. SOPHIE has been extensively used to characterize numerous planetary candidates discovered by \textit{TESS} \citep[e.g.,][]{bell2024toi,martioli2023toi, heidari2022hd,moutou2021toi}.

For this study, we obtained four spectra of TOI-6041 between August and October 2024. The observations were conducted in SOPHIE's high-resolution mode (R = 75,000) with simultaneous sky monitoring to account for potential contamination from moonlight. Exposure times ranged from 900 to 2100 seconds, depending on weather conditions, and were chosen to achieve a target S/N per pixel of approximately 50 at 550 nm. Two of the spectra had slightly lower S/N values of 45 and 39, but they were still sufficient for our analysis.

The RVs were derived using the standard SOPHIE pipeline, which computes cross-correlation functions (CCFs) as described by \cite{2009EAS....37..247B} and further refined by \cite{heidari2024sophie}. The resulting RV measurements presented in Table. \ref{tab:rvs SOPHIE} and have typical uncertainties of approximately 2 m/s and a mean dispersion of 11.9 m/s.

\subsection{TRES spectroscopy}

A reconnaissance spectrum of TOI-6041 was obtained with the Tillinghast Reflector Echelle Spectrograph \citep[TRES;][]{gaborthesis}, mounted on the 1.5 m Tillinghast Reflector telescope at the Fred Lawrence Whipple Observatory (FLWO) on Mount Hopkins, Arizona. TRES is an optical, fiber-fed echelle spectrograph covering a wavelength range of 390–910 nm with a resolving power of $R \approx 44,000$. The TRES spectrum was extracted following the procedure described by \cite{buchhave2010}. This spectrum is used only for our stellar classification presented in Sect. \ref{subsec:spectralanalysis}.

\subsection{High resolution imaging}
\label{high_reso_image}

As a usual part of the validation and confirmation process for transiting
exoplanets, high-resolution imaging is one of the critical assets required. The
presence of a close companion star, whether truly bound
or in the line of sight, will provide ``third-light” contamination of the observed transit, leading to derived properties for the exoplanet and host star that are incorrect \citep{ciardi2015understanding,furlan2017densities,furlan2020unresolved,2018AJ....155..244B}. In addition, it has been shown that the presence of a close companion dilutes small planet transits (<1.2 $R_e$) to the point of
non-detection \citep{lester2021speckle}. Given that nearly one-half of FGK stars are in
binary or multiple star systems \citep{matson2018stellar}, high-resolution imaging
provides crucial information toward our understanding of each discovered exoplanet
as well as more global information on exoplanetary formation, dynamics and evolution \citep{howell2021nasa}.

\begin{table}
\caption{\label{stellar parameters} Stellar properties of TOI-6041}
\centering
\resizebox{\columnwidth}{!}{%
\begin{tabular}{lll}
\hline
Identifiers: &  & \\
 & TIC 192415680 & \\
 & \textit{Gaia} DR3 432549871529588608 & \\
 & TOI-6041 & \\
   & &\\
\hline
Parameter & TOI-6041 & References\\
\hline
  & Parallax and coordinates & \\
Parallax (mas) & $13.76 \pm 0.02$ & \textit{Gaia} DR3\\
Distance (pc) & $72.3 \pm 0.1$ & \textit{Gaia} DR3\\
$\alpha$ (h m s) & $03:04:14.5$ & \textit{Gaia} DR3\\ 
$\delta$ (d m s) & $+43:33:28.6$ & \textit{Gaia} DR3 \\ 
  & &\\
 & Photometric properties & \\
   & &\\
B-V & $0.86 \pm 0.02$ & HIP\\
V (mag) & $9.84 \pm 0.03$ & HIP \\
\textit{Gaia} (mag) & $9.713 \pm 0.003$ & \textit{Gaia} DR3\\
$G_{\rm BP}$ (mag) & $8.605 \pm 0.003$ & \textit{Gaia} DR3 \\
$G_{\rm RP}$ (mag) & $7.540 \pm 0.001$ & \textit{Gaia} DR3\\ 
\textit{TESS} (mag) & $7.58 \pm 0.01$ & \textit{TESS} \\
J (mag) & $8.50 \pm 0.02$ & 2MASS \\
H (mag) & $8.13 \pm 0.02$ & 2MASS \\
$K_s$ (mag) & $8.06 \pm 0.01$ & 2MASS \\
$W_1$ (mag) & $8.01 \pm 0.02$ & WISE \\
$W_2$ (mag) & $8.07 \pm 0.02$ & WISE \\
$W_3$ (mag) & $8.02 \pm 0.02$ & WISE \\
$W_4$ (mag) & $8.0 \pm 0.2$ & WISE \\
  & &\\
 & Spectroscopic properties &\\
   & &\\
Spectral type & G7 & LAMOST$^{1}$\\
$\log R'_{\rm HK}$ & $-4.7 \pm 0.1$ & SOPHIE (Sect. \ref{subsec:spectralanalysis})\\
$v \sin i$ (km s$^{-1}$) & $3.4 \pm 1.0$ & SOPHIE (Sect. \ref{subsec:spectralanalysis})\\
                       & $3.0 \pm 0.5$ & TRES (Sect. \ref{subsec:spectralanalysis}) \\
$[{\rm Fe/H}]$ & $0.02 \pm 0.04$ & SOPHIE (Sect. \ref{subsec:spectralanalysis})\\
               & $0.05 \pm 0.08$ & TRES (Sect. \ref{subsec:spectralanalysis}) \\
$\log g$ (cgs) & $4.4 \pm 0.1$ & SOPHIE (Sect. \ref{subsec:spectralanalysis})\\
               & $4.5 \pm 0.1$ & TRES (Sect. \ref{subsec:spectralanalysis}) \\
Trigonometric $\log g$ (cgs) & $4.49 \pm 0.03$ & SOPHIE (Sect. \ref{subsec:spectralanalysis})\\                
$T_{\rm eff}$ (K) & $5334 \pm 66$ & SOPHIE (Sect. \ref{subsec:spectralanalysis}) \\
                  & $5417 \pm 50$ [130]$^{2}$ & TRES (Sect. \ref{subsec:spectralanalysis}) \\              
 & &\\ 
 & Bulk properties &\\
 & &\\
Mass ($M_{\odot}$) & $0.85 \pm 0.01$ & Sect. \ref{subsec:spectralanalysis} \\
                  & $0.89 \pm 0.04$ & Sect. \ref{EXOFAST}\\
Radius ($R_{\odot}$) & $0.86 \pm 0.02$ [0.04]$^{2}$ & Sect. \ref{subsec:spectralanalysis} \\
                      & $0.88 \pm 0.03$ & Sect. \ref{EXOFAST}\\
$P_{\rm rot}$ (d) & $23 \pm 5$ & Sect. \ref{subsec:spectralanalysis}\\
Age (Gyr) & $2.0 \pm 1.1$ & Gyrochronology (Sect. \ref{subsec:spectralanalysis})\\
          & $7.9^{+3.7}_{-3.9}$ & MIST model (Sect. \ref{EXOFAST})\\
\hline
\end{tabular}
}
\tablefoot{$^{1}$ \cite{zhang2022stellar}, $^{2}$ Adopted the systematic uncertainty floor suggested by \cite{tayar2022guide}.}
\end{table}

TOI-6041 was observed on 2020 December 02 and 2021 February 03 using the ‘Alopeke speckle instrument on the Gemini North 8-m telescope \citep{scott2021twin}.  ‘Alopeke provides simultaneous speckle imaging in two bands (562 nm and 832 nm) with output final data products including a reconstructed image with robust
contrast limits on companion detections \citep[e.g.,][]{howell2016speckle}. While both
observations had consistent results that TOI-6041 is a single star, within the achieved angular resolution and 5$\sigma$ magnitude contrast limits, the December 2020 observation had better seeing which led to deeper contrast levels in the 562 nm band. Four sets of 1000 X 0.06 second images were obtained and processed in our
standard reduction pipeline \citep[see][]{howell2011speckle}. Figure \ref{image} shows our final contrast curves and the 562 nm and 832 nm reconstructed speckle images. We find that TOI-6041 is a single star with no companion brighter than 5-9 magnitudes below that of the target star from the 8-m telescope diffraction limit (20 mas) out to 1.2”. At the distance of TOI-6041 (d=72 pc) these angular limits correspond to spatial limits of 1.44  to 86.4 AU.

\section{Stellar characterization}
\label{subsec:spectralanalysis}

To determine the stellar atmospheric parameters, we first combined the SOPHIE spectra of TOI-6041 to construct a high S/N template, after correcting for both the star's RV variation and the barycentric Earth RV. We then applied the methodology of \cite{sousa2011spectroscopic}, \cite{santos2013sweet} and \cite{sousa2018sweet} to derive the effective temperature ($T_{\rm eff}$), metallicity ([Fe/H]) and surface gravity ($\log g$). A more accurate trigonometric $\log g$ was also derived using {\it Gaia} DR3 data, following the procedure described in \citet{2021A&A...656A..53S}. Additionally, we estimated the projected stellar rotational velocity (v$\sin i$) for each individual SOPHIE spectrum using the calibration of \cite{boisse2010sophie}, then computed the mean value, which we report in Table \ref{stellar parameters} along with other stellar properties.

\begin{figure}
    \centering
    \includegraphics[width=\linewidth,trim=100 70 50 50,clip]{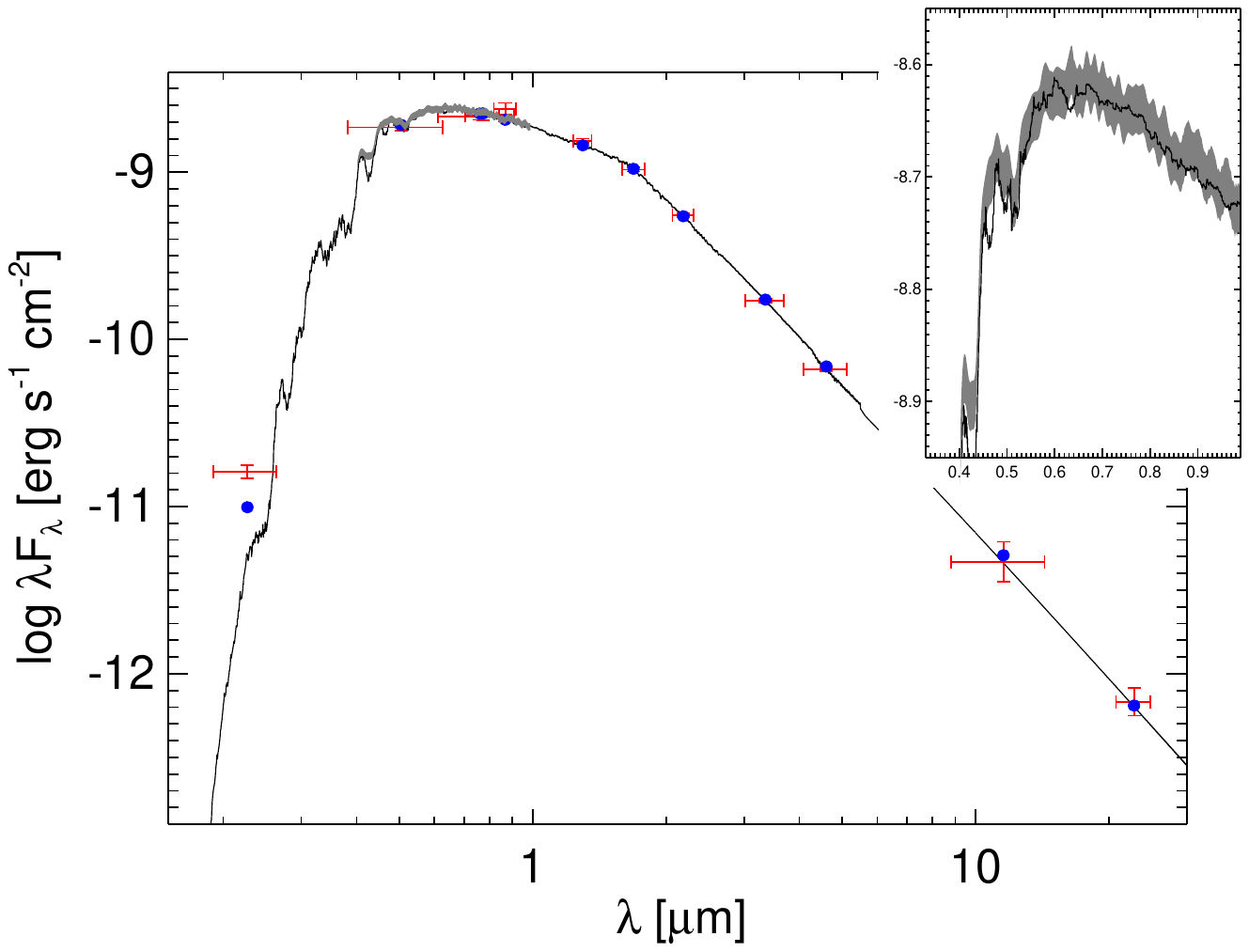}
    \caption{Spectral energy distribution of TOI-6041. Red symbols represent the observed photometric measurements, where the horizontal bars represent the effective width of the pass-band. Blue symbols are the model fluxes from the best-fit PHOENIX atmosphere model (black). The absolute flux-calibrated \textit{Gaia} spectrum is shown as a grey swathe in the inset figure.}
    \label{fig:sed}
\end{figure}

As an independent analysis, we also derived stellar parameters from the TRES spectrum using the Stellar Parameter Classification \citep[SPC;][]{buchhave2012} tool. SPC cross-correlates the observed spectrum with a grid of synthetic spectra based on Kurucz atmospheric models \citep{kurucz1992} to determine the star's $T_{\rm eff}$, [Fe/H], $\log g$, and v$\sin i$. Following the recommendation of \citet{tayar2022guide}, we conservatively adopted and reported in brackets in Table~\ref{stellar parameters} a systematic uncertainty floor on the TRES-derived $T_{\rm eff}$. For the SOPHIE-derived $T_{\rm eff}$, the systematic uncertainty had already been included in the reported uncertainty \citep{sousa2011spectroscopic}. The stellar parameters derived from the TRES spectrum are consistent within 1$\sigma$ with those obtained from the SOPHIE spectra, as summarized in Table~\ref{stellar parameters}.

An analysis of the broadband spectral energy distribution (SED) of the star was performed together with the \textit{Gaia} DR3 parallax, in order to determine an empirical measurement of the stellar radius \citep{Stassun:2016,Stassun:2017,Stassun:2018}. The $JHK_S$ magnitudes were sourced from {\it 2MASS} \citep{skrutskie2006two}, the W1--W4 magnitudes from {\it WISE} \citep{wright2010wide}, the $G_{\rm BP} G_{\rm RP}$ magnitudes from \textit{Gaia} \citep{gaia2023gaia}, and the NUV magnitude from {\it GALEX} \citep{martin2005galaxy}. The absolute flux-calibrated \textit{Gaia} spectrum was also utilized, when available. Together, the available photometry spans the full stellar SED over the wavelength range of at least 0.4--10~$\mu$m and as much as 0.2--20~$\mu$m (see Figure~\ref{fig:sed}).

A fit using PHOENIX stellar atmosphere models \citep{Husser:2013} was performed, adopting from the SOPHIE spectroscopic analysis the $T_{\rm eff}$, [Fe/H], and $\log g$. The extinction $A_V$ was fitted for, limited to the maximum line-of-sight value from the Galactic dust maps of \citet{Schlegel:1998}. 
The resulting fit (Figure~\ref{fig:sed}) has a $\chi^2 = 16.2$, excluding the {\it GALEX} NUV flux, which indicates a moderate level of activity \citep{findeisen2011stellar}, with best-fit $A_V = 0.03 \pm 0.03$. Integrating the (unreddened) model SED gives the bolometric flux at Earth, $F_{\rm bol} = 3.233 \pm 0.075 \times 10^{-9}$ erg~s$^{-1}$~cm$^{-2}$. Taking the $F_{\rm bol}$ together with the \textit{Gaia} parallax directly gives the bolometric luminosity, $L_{\rm bol} = 0.532 \pm 0.012$~L$_\odot$.  The Stefan-Boltzmann relation then gives the stellar radius, $R_\star = 0.855 \pm 0.023$~R$_\odot$. In addition, the stellar mass was estimated using the empirical relations of \citet{Torres:2010}, with the offset correction applied following \citet{santos2013sweet}, yielding $M_\star = 0.85 \pm 0.01$~M$_\odot$. We note that the intrinsic error from the \citet{Torres:2010} calibration was also added in quadrature to the final uncertainty.

The formal error estimates for the stellar radius from our SED analysis may be underestimated. For a star like TOI-6041, the systematic uncertainty floor is expected to be approximately 4.2\% in radius, as suggested by \citet{tayar2022guide}. To account for this, we conservatively adopt these relative uncertainties and report them in brackets in Table \ref{stellar parameters}. In the case of the stellar mass, our uncertainty is consistent with this systematic uncertainty floor.

We estimate the stellar rotation period using the method described in \citet{mascareno2016magnetic}, which relies on the activity--rotation relation calibrated for different spectral types using the chromospheric activity index $\log~ (R'_{\rm HK})$. For this G-type star, we applied the corresponding calibration with $\log ~(R'_{\rm HK}) = -4.7 \pm 0.1$, derived from the SOPHIE spectra following the procedures of \citet{noyes1984rotation} and \citet{boisse2010sophie}. We compare this with the value obtained from \citet{Mamajek:2008}, which incorporates both $\log(R'_{\mathrm{HK}})$ and the $B\!-\!V$ color. These methods yield rotation periods of $23 \pm 5$ days and $29 \pm 5$ days, respectively, consistent within approximately $1\sigma$. 

To further investigate the stellar rotation period, we applied the Lomb-Scargle periodogram to the SAP TESS light curves from all three available sectors, with periods from 0 to 50 d. We detected a strong, broad peak between approximately 20 and 30 d, with a false alarm probability below 0.001\% (see the periodogram and phase-folded light curve in Fig.~\ref{rotation_period}). This signal is consistent with the rotation periods estimated from $\log(R'_{\mathrm{HK}})$ and is likely associated with rotational modulation. Its broad nature may be attributed to the limited time baseline of each TESS sector, which spans only about 27 d. 

Additionally, using the empirical age–activity relations of \citet{Mamajek:2008}, we estimate an age of $2.0 \pm 1.1$~Gyr.

Finally, as a further check, we compared these values with results obtained using an alternative modeling method in \texttt{EXOFASTv2}, performed simultaneously with the planetary fit (see Sect.~\ref{EXOFAST}). The parameters agree within 1$\sigma$, except for the stellar age, which is consistent only at the 2$\sigma$ level. However, we note that the stellar age derived from the \texttt{EXOFASTv2} fit is poorly constrained, yielding a broad and asymmetric posterior distribution (see Fig.~\ref{age}).

\section{Detection and characterization of TOI-6041's planets}
\label{detection}

\subsection{RV-only analysis}
\label{rv_only}

We began our RV analysis by performing a periodogram analysis of the RVs and the S-index activity indicator using the Data and Analysis Center for Exoplanets \citep[DACE,][]{delisle2016analytical}\footnote{Available at \url{https://dace.unige.ch}}. Since we have only four SOPHIE RV measurements, we use only the APF data in this section to avoid instrumental offsets between the two datasets, while we also tested fits including the both datasets in our global modeling in Sect. \ref{EXOFAST}.

\begin{figure}
\centering
\includegraphics[width=0.9\columnwidth]{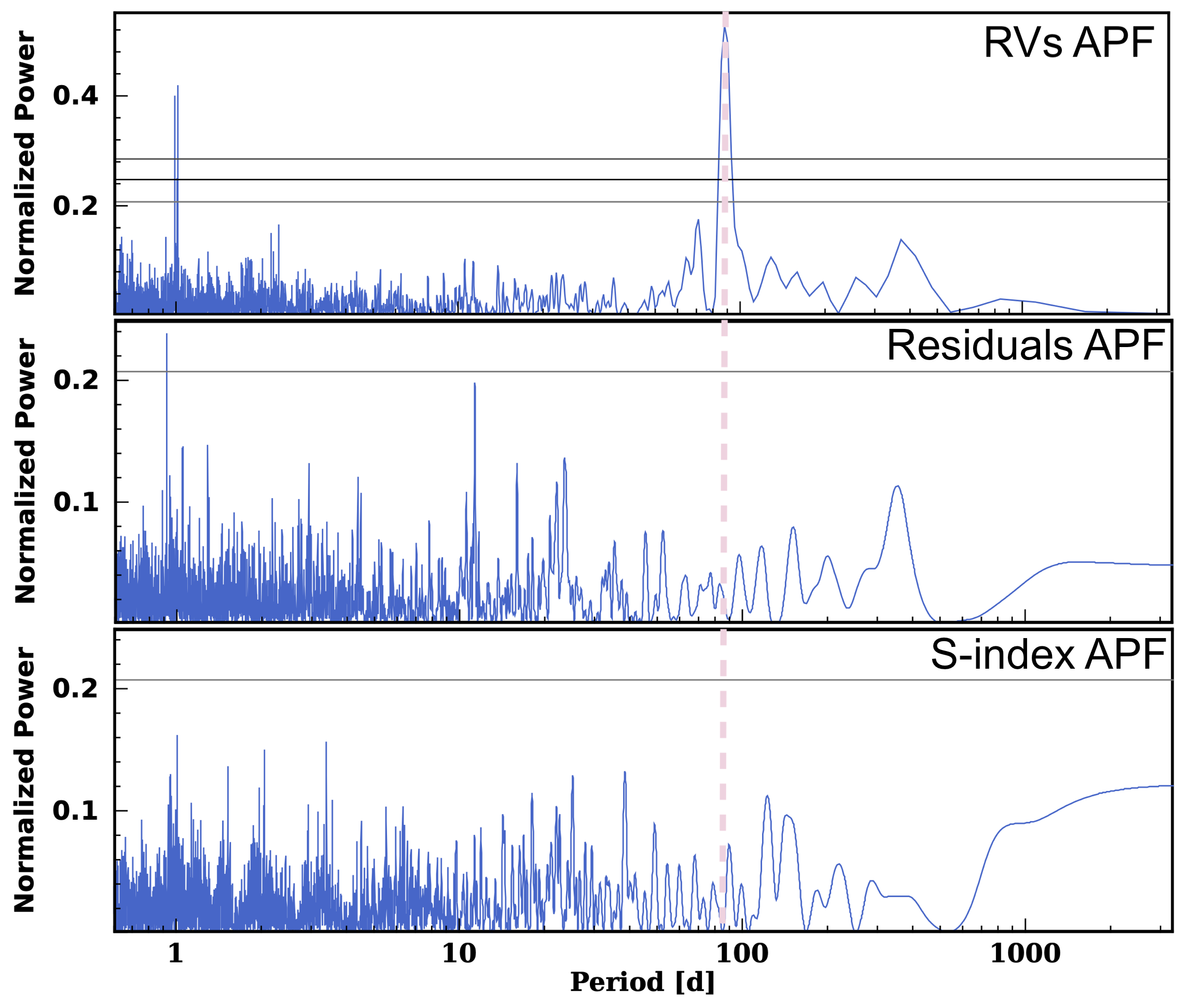}
\caption{Periodograms of RV and S-index derived from APF data for TOI-6041. From \emph{top} to \emph{bottom}: the APF RVs, the residuals of the RVs after a Keplerian fit to the 88 d signal, and the S-index activity indicator. The vertical pink line marks the planet candidate with a period of 88 d, which shows no corresponding signal in the S-index periodogram. The horizontal lines indicate false alarm probability \citep[FAP,][]{2008MNRAS.385.1279B} thresholds of 0.1\%, 1\%, and 10\%, respectively from top to bottom.}
\label{periodogram}
\end{figure}

The periodogram of RVs showed a significant peak at 88 d, below the analytical false alarm probability \citep[FAP,][]{2008MNRAS.385.1279B} of 0.1\% (see Fig. \ref{periodogram}, first panel). We modelled and subtracted this signal using a circular Keplerian fit (see justification below). The periodogram of the residuals then reveals a peak at 0.9\,d and a tentative signal around 11\,d, both with FAP values close to 10\% (see Fig.~\ref{periodogram}, second panel). These two signals are aliases of each other and are not statistically significant. Additional RV measurements are required to understand their origin. Finally, we detect neither a significant signal in the S-index activity periodogram (see Fig. \ref{periodogram}, third panel) nor any correlation between the S-index and the RVs (Pearson's correlation coefficient r =$-0.1$) or their residuals (r=0.3) after removing the 88 d signal.

\begin{table}
\centering
\begin{tabular}{lc}
\hline
Model & $\Delta \ln Z$ \\
\hline
No-planet & 0 \\
1cpl & 25.0 \\
1pl & 25.3 \\
2cpl  & 25.0 \\
3cpl & 25.7 \\
\hline
\end{tabular}
\caption{A comparison of different models tested using RV-only data analyzed with \texttt{juliet}. The tested models include: 1) No-planet model, 2) a one-planet model with an 88 d signal and free eccentricity (1pl), 3) a one-planet model with a circular orbit (1cpl), 4) a two-planet model with circular orbits at 88 d and 26 d (2cpl), where the period and $T_0$ for the 26 d planet are fixed, and 5) a three-planet model, which includes the same setup as the 2cpl model for the planets at 26 d and 88 d, with an additional planet on a circular orbit and a uniform prior on the period ranging from 0 to 20 d (3cpl).}
\label{rv-model}
\end{table}

\begin{figure}
\centering
\includegraphics[width=0.8\columnwidth]{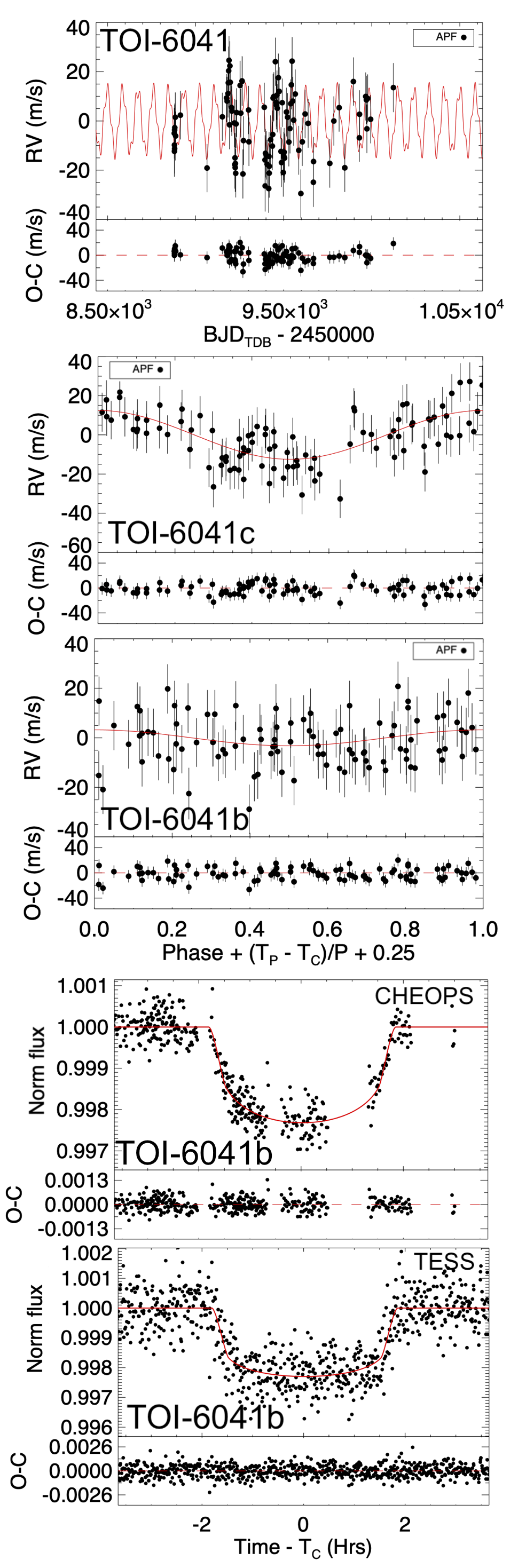}
\caption{APF RV measurements of TOI-6041 (\emph{top panel}); phase-folded RVs for TOI-6041c (\emph{second panel}) and TOI-6041b (\emph{third panel}); and the phase-folded \textit{CHEOPS} and \textit{TESS} light curves for TOI-6041b (\emph{fourth and fifth panels}). The red lines show the best-fit models derived using EXOFASTv2. Residuals are displayed at the bottom of each panel.}
\label{6041_rvs_photometry}
\end{figure}

Additionally, we considered several RV-only models and performed model comparisons. The RV analysis was conducted using \texttt{juliet} \citep{espinoza2019juliet}, which employs \emph{RadVel} \citep{fulton2018radvel} to model the RVs. Since we found no significant influence of stellar activity in our RVs, we did not include an Gaussian process in our analysis. For each tested model, \texttt{juliet} computes the Bayesian log evidence (\(\ln Z\)). A model is considered moderately favored over another if the difference in Bayesian log evidence (\(\Delta\ln Z\)) exceeds two, while a difference greater than five indicates strong preference \citep{trotta2008bayes}. When \(\Delta\ln Z \leq 2\), the models are statistically indistinguishable, and in such cases, the model with the fewest free parameters is preferred.

First, we tested a no-planet model (No-planet) and a one-Keplerian model with  circular orbit and a uniform prior on the orbital period between 70 and 100 d (1cpl). For the remaining parameters, we adopted fairly broad uniform priors. The one-Keplerian model converged on the 88 d signal. As shown in Table \ref{rv-model}, the \(\Delta\ln Z\) between these two models is 25.0, indicating a strong detection of the 88 d signal. We then considered a one-Keplerian model with a freely varying eccentricity (1cpl), which yielded a best-fit eccentricity of \( e = 0.3 \pm 0.2 \). However, since the eccentricity is not significantly detected and the \(\Delta\ln Z\) values for the circular and eccentric models are indistinguishable, we adopted the circular model for our final analysis.

Next, we tested a two-Keplerian model on 88 d and 26 d signals (2cpl). Since we did not observe a signature for the 26 d periodic signal in the RV periodogram, we fixed the transit epoch ($T_0$) and period to the values derived from photometry for this planet (see Sect. \ref{sec:CHEOPS_photometry}) and ran the model. This model converged with an amplitude of $K = 2.6^{+1.3}_{-1.2}$ for this signal, but it reached the prior edge. The $\Delta\ln Z$ comparison showed this model is indistinguishable from the one-Keplerian model, indicating no significant detection of 26 d signal in RVs, consistent with the results of the periodogram.

Finally, we tested a three-Keplerian model with an additional planet candidate with a circular orbit and a uniform prior on the period ranging from 0 to 20 d (3cpl). This model was motivated by the tentative 0.9 d and 11 d signals observed in the residual periodogram. The fit converged on an 11 d signal. However, this model remained statistically indistinguishable from the one-planet model, confirming that the 11 d signal is not statistically significantly detected.

Therefore, we conclude that the only robustly detected signal in our RV data is the 88 d signal. Since this signal has no corresponding signal in the S-index activity indicator, it is likely of planetary origin. For the rest of this article, we refer to it as TOI-6041 c.

\subsection{Global modeling}
\label{EXOFAST}
To simultaneously model the stellar and planetary parameters, we used \texttt{EXOFASTv2} \citep{Eastman2013, Eastman2017, eastman2019}. Our analysis incorporated photometric data from three \textit{TESS} sectors, three \textit{CHEOPS} observations, and RVs from APF. We applied Gaussian priors on [Fe/H], centered on the value derived from the SOPHIE spectrum (Sect.~\ref{subsec:spectralanalysis}), but adopted a conservatively broad standard deviation equal to twice its reported uncertainty. A Gaussian prior was also applied to the parallax, using values from the \textit{Gaia} DR3 catalog. Additionally, we imposed an upper limit on extinction, $A_V$, based on the dust maps from \citet{Schlegel:1998}. The broadband photometry, as presented in Table~\ref{stellar parameters}, is also included. The \textit{Gaia} DR3 parallax, the $A_V$ upper limit, and the broadband

\onecolumn
\providecommand{\bjdtdb}{\ensuremath{\mathrm{BJD}_{\mathrm{TDB}}}}
\providecommand{\feh}{\ensuremath{\left[\mathrm{Fe}/\mathrm{H}\right]}}
\providecommand{\teff}{\ensuremath{T_{\mathrm{eff}}}}
\providecommand{\teq}{\ensuremath{T_{\mathrm{eq}}}}
\providecommand{\ecosw}{\ensuremath{e\cos{\omega_*}}}
\providecommand{\esinw}{\ensuremath{e\sin{\omega_*}}}
\providecommand{\msun}{\ensuremath{\,M_\odot}}
\providecommand{\rsun}{\ensuremath{\,R_\odot}}
\providecommand{\lsun}{\ensuremath{\,L_\odot}}
\providecommand{\mj}{\ensuremath{\,M_{\mathrm{J}}}}
\providecommand{\rj}{\ensuremath{\,R_{\mathrm{J}}}}
\providecommand{\me}{\ensuremath{\,M_{\mathrm{E}}}}
\providecommand{\re}{\ensuremath{\,R_{\mathrm{E}}}}
\providecommand{\fave}{\ensuremath{\langle F \rangle}}

\providecommand{\fluxcgs}{10$^9$ erg s$^{-1}$ cm$^{-2}$}

\begin{deluxetable}{lcccc}
\tablewidth{0.8\textwidth}
\tabletypesize{\scriptsize}
\tablecaption{Median values and 68\% confidence intervals for TOI-6041 and its two planets.}
\tablehead{
  \colhead{Parameter} & 
  \colhead{Description} & 
  \colhead{Model Priors} & 
  \multicolumn{2}{c}{Value} \\

}
\startdata

\multicolumn{5}{l}{Stellar Parameters:} \\
~~~~$M_*$\dotfill & Mass ($M_\odot$)\dotfill &--& $0.885^{+0.042}_{-0.036}$\\
~~~~$R_*$\dotfill & Radius ($R_\odot$)\dotfill &--& $0.879\pm0.026$ \\
~~~~$L_*$\dotfill & Luminosity ($L_\odot$)\dotfill &--& $0.613\pm0.033$ \\
~~~~$F_{Bol}$\dotfill &Bolometric Flux (cgs)\dotfill &--&$3.73\pm0.20 \times 10^{-9}$\\
~~~~$\rho_*$\dotfill &Density (cgs)\dotfill & --&$1.84^{+0.19}_{-0.17}$\\
~~~~$\log{g}$\dotfill &Surface gravity (cgs)\dotfill&-- &$4.498^{+0.031}_{-0.030}$\\
~~~~$T_{\rm eff}$\dotfill &Effective temperature (K)\dotfill&-- &$5445^{+97}_{-96}$\\
~~~~$[{\rm Fe/H}]$\dotfill &Metallicity (dex)\dotfill &$\mathcal{N}$[0.02, 0.08]&$0.104^{+0.069}_{-0.068}$\\
~~~~$[{\rm Fe/H}]_{0}$\dotfill &Initial Metallicity$^{1}$ \dotfill &--&$0.111^{+0.069}_{-0.067}$\\
~~~~$Age$\dotfill &Age (Gyr)\dotfill &--&$7.9^{+3.7}_{-3.9}$\\
~~~~$EEP$\dotfill &Equal Evolutionary Phase$^{3}$ \dotfill &--&$359^{+27}_{-21}$\\
~~~~$A_V$\dotfill &V-band extinction (mag)\dotfill &$\mathcal{U}$[0.01, 0.74]&$0.260^{+0.062}_{-0.070}$\\
~~~~$\sigma_{SED}$\dotfill &SED photometry error scaling \dotfill &--&$0.36^{+0.17}_{-0.10}$\\
~~~~$\varpi$\dotfill &Parallax (mas)\dotfill &$\mathcal{N}$[13.79, 0.02]&$13.794\pm0.020$\\
~~~~$d$\dotfill &Distance (pc)\dotfill &--&$72.49\pm0.10$\\
\\
Planetary Parameters:&&&TOI-6041\,b&TOI-6041\,c\\
~~~~$P$\dotfill &Period (d)\dotfill &--&$26.04945^{+0.00033}_{-0.00034}$&$88.0^{+1.6}_{-1.3}$\\
~~~~$R_P$\dotfill &Radius ($R_\oplus$)\dotfill &--&$4.55^{+0.18}_{-0.17}$&---\\
~~~~$M_p$\dotfill &Mass ($M_\oplus$)\dotfill &--&$10.81\pm 6.36$ ($<$ 28.9$^{2}$)&---\\
~~~~$T_C$\dotfill &Observed Time of conjunction$^{4}$ ((BJD$_{\mathrm{TDB}}$)
\dotfill &--&$2459890.1062^{+0.0086}_{-0.0085}$&--\\
~~~~$a$\dotfill &Semi-major axis (AU)\dotfill &--&$0.1651^{+0.0026}_{-0.0022}$&$0.3721^{+0.0067}_{-0.0063}$\\
~~~~$i$\dotfill &Inclination (Degrees)\dotfill &--&$88.955^{+0.063}_{-0.062}$&$90.00$~(fixed)\\
~~~~$\dot{\omega}_{\rm GR}$\dotfill &Computed GR precession ($^\circ$/century)\dotfill &--&$0.0801^{+0.0025}_{-0.0022}$&$0.01054^{+0.00041}_{-0.00039}$\\
~~~~$T_{\rm eq}$\dotfill &Equilibrium temp$^{5}$ (K)\dotfill &--&$605.6^{+8.4}_{-8.3}$&$403.5\pm5.9$\\
~~~~$\tau_{\rm circ}$\dotfill &Tidal circ timescale (Gyr)\dotfill &--&$11500^{+7800}_{-6800}$&$2.7^{+3.5}_{-1.4} \times 10^{5}$\\
~~~~$K$\dotfill &RV semi-amplitude (m/s)\dotfill &--&$2.5\pm1.5$~($<$ 6.7 $^{2}$)&$12.1\pm1.4$\\
~~~~$R_P/R_*$\dotfill &Radius of planet in stellar radii \dotfill &--&$0.04753^{+0.00058}_{-0.00057}$&--\\
~~~~$a/R_*$\dotfill &Semi-major axis in stellar radii \dotfill &--&$40.4\pm1.3$&--\\
~~~~$\delta$\dotfill &$\left(R_P/R_*\right)^2$ \dotfill &--&$0.002259^{+0.000056}_{-0.000054}$&--\\
~~~~$\delta_{\rm CHEOPS}$\dotfill &Transit depth in \textit{CHEOPS} (frac)\dotfill &--&$0.002360^{+0.000040}_{-0.000039}$&--\\
~~~~$\delta_{\rm TESS}$\dotfill &Transit depth in \textit{TESS} (frac)\dotfill &--&$0.002321\pm0.000038$&--\\
~~~~$\tau$\dotfill &In/egress transit duration (d)\dotfill &--&$0.0145\pm0.0011$&--\\
~~~~$T_{14}$\dotfill &Total transit duration (d)\dotfill &--&$0.1526\pm0.0013$&--\\
~~~~$T_{FWHM}$\dotfill &FWHM transit duration (d)\dotfill &--&$0.1381\pm0.0011$&--\\
~~~~$b$\dotfill &Transit impact parameter \dotfill &&$0.738^{+0.020}_{-0.022}$&--\\
~~~~$\delta_{S,2.5\mu m}$\dotfill &BB eclipse depth at 2.5$\mu$m (ppm)\dotfill &--&$0.317^{+0.043}_{-0.039}$&--\\
~~~~$\delta_{S,5.0\mu m}$\dotfill &BB eclipse depth at 5.0$\mu$m (ppm)\dotfill &--&$13.71^{+1.0}_{-0.96}$&--\\
~~~~$\delta_{S,7.5\mu m}$\dotfill &BB eclipse depth at 7.5$\mu$m (ppm)\dotfill &--&$41.9^{+2.4}_{-2.3}$&--\\
~~~~$\rho_P$\dotfill &Density (cgs)\dotfill &--&$0.62^{+0.39}_{-0.36}$ ($<$ 1.79$^{2}$)&--\\
~~~~$logg_P$\dotfill &Surface gravity (cgs)\dotfill &--&$2.72^{+0.20}_{-0.32}$&$2.85^{+0.14}_{-0.12}$\\
~~~~$\Theta$\dotfill &Safronov Number \dotfill &--&$0.031\pm0.018$&$0.221^{+0.040}_{-0.030}$\\
~~~~$\fave$\dotfill &Incident Flux (\fluxcgs)\dotfill &--&$0.0305^{+0.0017}_{-0.0016}$&$0.00601^{+0.00036}_{-0.00034}$\\
~~~~$T_S$\dotfill &Observed Time of eclipse$^{4}$ (BJD$_{\mathrm{TDB}}$)\dotfill &--&$2459903.1291^{+0.0086}_{-0.0085}$&--\\
~~~~$T_P$\dotfill &Time of Periastron (TJD$_{\mathrm{TDB}}$)\dotfill &--&$2459890.1053^{+0.0086}_{-0.0085}$&$2.458774^{+0.000010}_{-0.000013} \times 10^{6}$\\
~~~~$T_A$\dotfill &Time of asc node (TJD$_{\mathrm{TDB}}$)\dotfill &--&$2459909.6424^{+0.0086}_{-0.0085}$&$2458839.6^{+9.1}_{-12}$\\
~~~~$T_D$\dotfill &Time of desc node (TJD$_{\mathrm{TDB}}$)\dotfill &--&$2459896.6176^{+0.0086}_{-0.0085}$&$2458795.6^{+9.8}_{-12}$\\
~~~~$M_P\sin i$\dotfill &Minimum mass ($M_\oplus$)\dotfill &--&$10.81\pm 6.36$ ($<$ 28.9$^{2}$) &$77.9^{+9.2}_{-8.9}$\\
~~~~$d/R_*$\dotfill &Separation at mid transit \dotfill &--&$40.4\pm1.3$&--\\
~~~~$P_T$\dotfill &A priori non-grazing transit prob \dotfill &--&$0.02356^{+0.00074}_{-0.00073}$&--\\
\enddata
\label{toi6041_result_exofast}
\tablefoot{$^{1}$ The metallicity of the star at birth; $^{2}$ The 3$\sigma$ upper limits; $^{3}$ Corresponds to static points in a star's evolutionary history. See Section. 2 in \cite{dotter2016mesa}, $^{4}$ Reference epoch = 2456438.359500; $^{5}$ Assumes no albedo and perfect redistribution }
\end{deluxetable}

\begin{table}
\centering
\caption{\centering Median values and 68\% confidence interval for transit times, and TTVs.}
\begin{tabular}{lcccc}
Epoch & Observed mid-transit (BJD) & TTV (min) \\
\hline
$-42$ (\textit{TESS}) & $2458796.0279^{+0.0014}_{-0.0015}$ & $-5.8 \pm 2.1$ \\
0 (\textit{TESS}) & $2459890.0860 \pm 0.0012$ & $-40.9 \pm 1.7$ \\
14 (\textit{CHEOPS}) & $2460254.81990 \pm 0.00049$ & $16.28 \pm 0.70$ \\
16 (\textit{CHEOPS}) & $2460306.9173^{+0.0038}_{-0.0047}$ & $13.7 \pm 6.1$ \\
28 (\textit{TESS})& $2460619.4775 \pm 0.0011$ & $-36.4 \pm 1.6$ \\
29 (\textit{CHEOPS})& $2460645.5292^{+0.0017}_{-0.0022}$ & $-33.3 \pm 2.8$ \\
\hline
\end{tabular}
\label{ttv_table}
\end{table}

\twocolumn

\noindent  photometry were used to constrain the stellar properties through SED fitting. To compare our results, we also modeled the star using the MESA Isochrones and Stellar Tracks \citep[MIST;][]{dotter2016mesa, choi2016mesa}. As part of the modeling process, we simultaneously detrended the \textit{TESS} photometry using a spline \citep{vanderburg2014technique} within \texttt{EXOFASTv2} to mitigate instrumental systematics (see Sect.~\ref{sec:CHEOPS_photometry} for the \textit{CHEOPS} detrending procedure). We did not impose any priors on the planetary parameters, allowing \texttt{EXOFASTv2} to explore them freely (see \cite{Eastman2013, Eastman2017, eastman2019} for detailed information).

We began our analysis with a circular orbit model for both planets. Upon convergence, we observed a significant TTV signal for TOI-6041 b (see Fig. \ref{ttv}). To account for this, we incorporated a TTV model using \texttt{EXOFASTv2} as part of our joint analysis and re-ran the fit. We note that the TTVs in this model correspond to artificial shifts added to a Keplerian model, which allows for a first estimate of the system parameters. In most cases, such an approach provides suitable parameter estimates \citep[see, e.g.,][]{hebrard2020discovery,almenara2022photodynamical}. Mutual gravitational interactions are taken into account below in Section~\ref{dynamics}. The updated model, which included TTVs, was strongly favored by the Bayesian Information Criterion (BIC), with a difference of $\Delta \mathrm{BIC} = 543.8$ compared to the Keplerian-only model. This indicates a significantly better fit to the data. We chose this model as the final one and present the posterior distributions of all model parameters in Table~\ref{toi6041_result_exofast} and Table~\ref{Wavelength_exo}, while the best-fit model for both the RV and photometric data is shown in Fig.~\ref{6041_rvs_photometry}.

Including TTVs in the model allowed the transit times to deviate from those predicted by a purely Keplerian orbit and a strictly linear ephemeris. \texttt{EXOFASTv2} determined the mid-transit times based on both a linear ephemeris and a TTV model, and provide the corresponding observed-minus-calculated (O–C) diagram (see Fig.~\ref{ttv_o_c}). A summary of the observed mid-transit time and the TTV measurements is provided in Table~\ref{ttv_table} (see Sect.~\ref{dynamics} for a more detailed discussion of the system's TTVs). The TTVs display a peak-to-peak variation of 57.2 minutes and an RMS scatter of 28 minutes.

We also tested fits that included the SOPHIE RVs. However, given that only four SOPHIE measurements were available, the corresponding RV jitter term could not be robustly constrained. Importantly, including or excluding these four data points did not affect our results. We therefore chose not to include the SOPHIE RVs in our final analysis. The results indicate that TOI-6041 has a mass of $0.885^{+0.042}_{-0.036}~M_\odot$, a radius of $0.879 \pm 0.026~R_\odot$, and an effective temperature of $5445^{+97}_{-96}$~K. These values are consistent within 1$\sigma$ with those derived from the spectral analysis and SED fitting presented in Sect.~\ref{subsec:spectralanalysis}, with the exception of the stellar age, which differs at the 2$\sigma$ level. However, we note that the age estimate from the \texttt{EXOFASTv2} analysis remains weakly constrained, yielding a broad and skewed posterior distribution (see Fig.\ref{age}).

For TOI-6041\,b, accounting for the observed TTV signal, we determine an average orbital period of $26.04945^{+0.00033}_{-0.00034}$\,d and a radius of $4.55^{+0.18}_{-0.17}\,R_\oplus$. Although this planet is not unambiguously detected in our RV data, the measurements allow us to place a 3$\sigma$ upper limit on its RV semi-amplitude of 7\,m\,s$^{-1}$, corresponding to a maximum mass of 28.9\,$M_\oplus$. Additional RV observations are needed to more tightly constrain the mass of TOI-6041\,b.

We also tested a model allowing for a non-zero eccentricity for TOI-6041\,b. However, the MCMC chains did not converge well due to a strong bimodality in the argument of periastron ($\omega$), with two dominant modes around $20^\circ$ and $230^\circ$. The posterior distribution of the eccentricity spans the entire prior range (0--1), with 68\% of the probability lying below 0.6. We therefore conclude that the current data do not support a significant eccentricity detection for TOI-6041\,b. As before, additional observations—both RV and TTV—are needed to better constrain the planet’s eccentricity.

The outer planet, TOI-6041\,c, is detected solely through our RV measurements. It has an orbital period of $88.0^{+1.6}_{-1.3}$ d and a minimum mass of $0.245^{+0.029}_{-0.028}~M_\mathrm{J}$. We adopted a circular orbit for this planet, as the eccentricity is not significantly constrained by the data (see Sect.~\ref{rv_only}). We also note that including or excluding an RV model for TOI-6041\,b does not affect the derived parameters of TOI-6041\,c. Whether TOI-6041\,c is transiting remains uncertain, as the time of periastron listed in Table~\ref{toi6041_result_exofast} has an uncertainty of approximately 12 d. Each TESS sector provides about 27 d of coverage, and a data gap occurs about near the midpoint of the relevant sector. It is therefore possible that a potential transit fell within this gap. Additional RV measurements are required to better constrain the time of periastron and to assess the planet’s transit probability.

\section{Dynamical analysis}
\label{dynamics}
In this section, we present a dynamical analysis of the TOI-6041 system to assess the long-term stability of its orbital configuration and to investigate whether the observed TTVs of planet b can be explained by its gravitational interactions with the outer planet c. Although several transits of planet b have been observed, the planet is not detected in RVs; planet c, conversely, is detected via RVs but does not transit. This limits our ability to constrain the system's orbital architecture. The analysis presented here should therefore be regarded as exploratory, pending additional observational constraints. Additional data would benefit from a
photodynamical analysis \citep[see, e.g.,][]{almenara2022photodynamical}.
\begin{figure*}
    \centering
    \includegraphics[width=\linewidth]{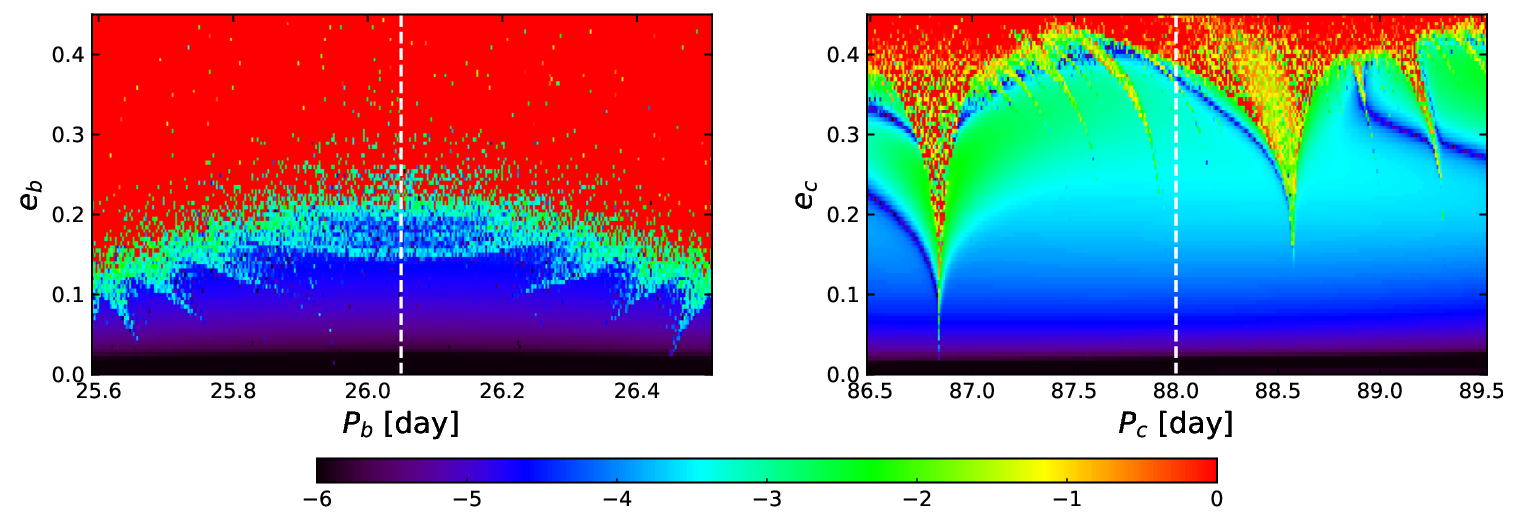}
    \caption{Stability analysis of the TOI-6041 system. The phase space of the system is explored by varying the orbital period $P_i$ and eccentricity $e_i$ of each planet independently. For each initial condition (Table~\ref{toi6041_result_exofast}), the system is integrated over 10 kyr, and a stability criterion is derived with the frequency analysis of the mean longitude. The chaotic diffusion is measured by the variation in the frequencies. The color scale corresponds to the logarithmic variation in the stability index used in \citep{Correia2010}. The red zone corresponds to highly unstable orbits, while the dark blue region can be assumed to be stable on a billion-year timescale. White dashed lines indicate the nominal orbital periods reported in Table~\ref{toi6041_result_exofast}.}
    \label{fig:stability_ecc}
\end{figure*}

\subsection{Stability Analysis}
\label{stability}
We first performed a global frequency analysis \citep{Laskar1990,LASKAR1993} in the vicinity of the best fitting circular solution (Table \ref{toi6041_result_exofast}), using the methodology described in \cite{Correia2005,Correia2010,Couetdic2010}. The goal is to assess the long-term stability of the orbital solution. The system was integrated on a regular 2D mesh of initial conditions, varying the orbital period and eccentricity of each planet individually, while keeping all other orbital elements fixed to the values listed in Table~\ref{toi6041_result_exofast}. The integrations were carried out using the symplectic integrator SABA1064 of \cite{Farres2013}, with a time step of $5 \times 10^{-3}$ yr, and included general relativity corrections. Each initial condition was integrated over $10$ kyr, and a stability indicator was derived with the frequency analysis of the mean longitude, to be the variation in the measured mean motion over the two consecutive $5$ kyr intervals of time \citep[for more details, see][]{Couetdic2010}. For regular motion, there is no significant variation in the mean motion along the trajectory, while it can vary significantly for chaotic trajectories. The results are shown in Figure \ref{fig:stability_ecc}, where “red” represents the strongly chaotic trajectories, and “dark blue” denotes extremely stable regions. We observe that the system remains stable for eccentricities up to $\sim 0.15$ for planet b and $\sim 0.3$ for planet c.

\begin{figure}
    \centering
    \includegraphics[width=\columnwidth]{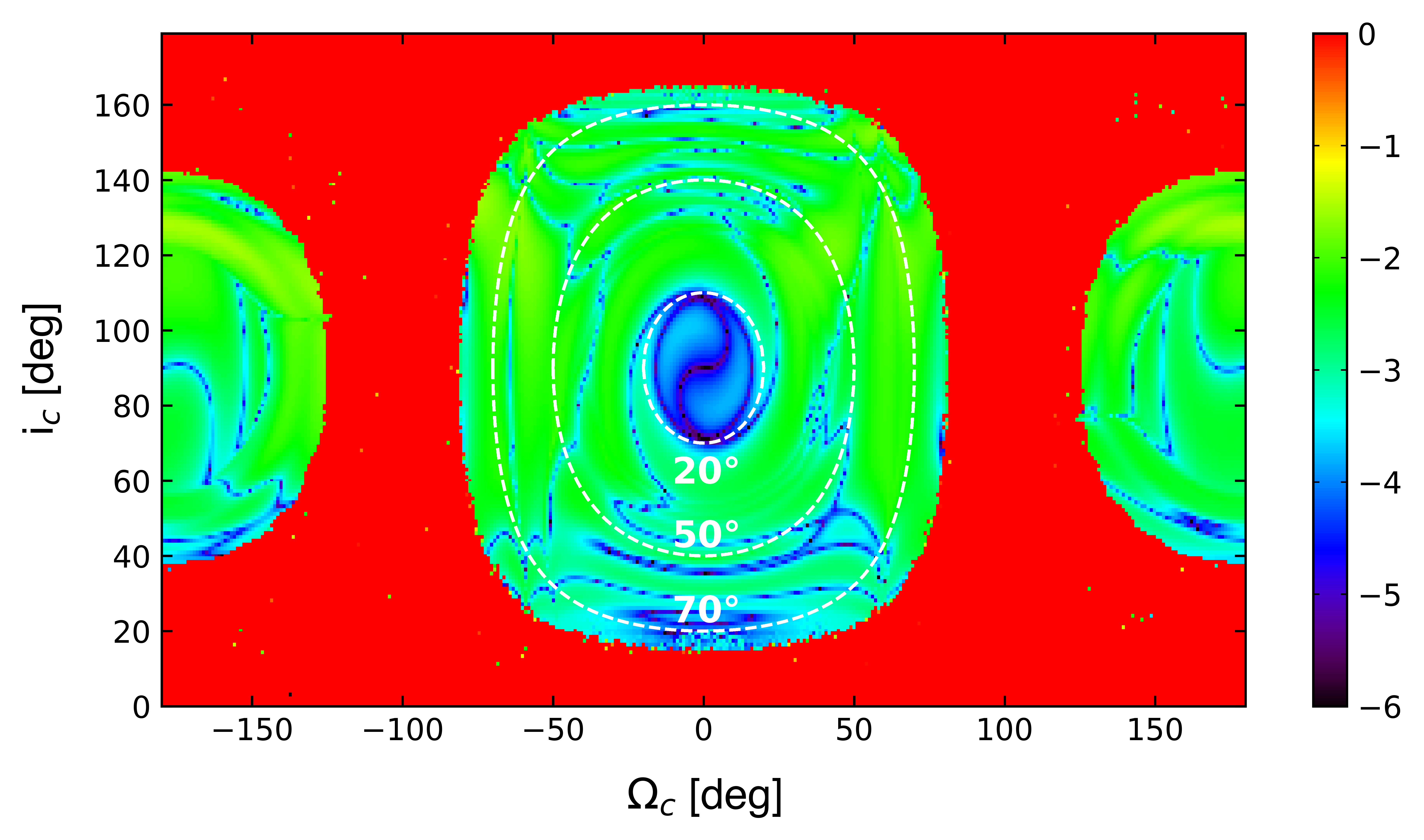}
    \caption{Stability analysis of the TOI-6041 system in the $(\Omega_\mathrm{c}, i_\mathrm{c})$ plane. The phase space is explored by varying these two parameters while all other orbital elements are fixed to the nominal values in Table~\ref{toi6041_result_exofast}. As in Figure~\ref{fig:stability_ecc}, red regions indicate strongly chaotic orbits, while dark blue regions correspond to long-term stability. White curves denote levels of constant mutual inclination between planets b and c.}
    \label{fig:stability_inc}
\end{figure}

Additionally, we computed a 2D stability map for the two unconstrained orbital parameters of planet c — the inclination $i_\mathrm{c}$ and the longitude of the ascending node $\Omega_\mathrm{c}$ — to assess how dynamical arguments can constrain their possible values. Figure~\ref{fig:stability_inc} explores the stability in the $(\Omega_\mathrm{c}, i_\mathrm{c})$ plane, with all other orbital elements fixed to the values listed in Table~\ref{toi6041_result_exofast}. We find that the system remains stable for $20^\circ \lesssim i_\mathrm{c} \lesssim 160^\circ$, corresponding to mutual inclinations up to about $70^\circ$. This dynamical constraint allows us to place an upper limit on the true mass of planet c, despite the absence of a transit. Given the 1$\sigma$ upper bound on the minimum mass $m_\mathrm{c} \sin i_\mathrm{c} < 0.274 M_{\text{Jup}}$, and the inclination $ i_\mathrm{c} \lesssim 160 ^\circ$, we infer a 1$\sigma$ upper limit on the true mass of $m_\mathrm{c} \lesssim 0.80 M_{\text{Jup}}$.

\subsection{Transit-timing Variations}
To investigate the origin of the observed TTVs in planet b, we generated synthetic TTV signals using \texttt{TTVFast} \citep{Deck2014} for different orbital configurations of planet c.  Adopting the circular and coplanar setup from Table~\ref{toi6041_result_exofast}, we found that the resulting TTV amplitudes remained below 2 minutes—significantly smaller than the observed variations. Reducing the inclination of planet~c (and thereby increasing its true mass) modestly increased the amplitude, but remained insufficient to match observations.

\begin{figure}[t!]
    \centering
    \includegraphics[width=\columnwidth]{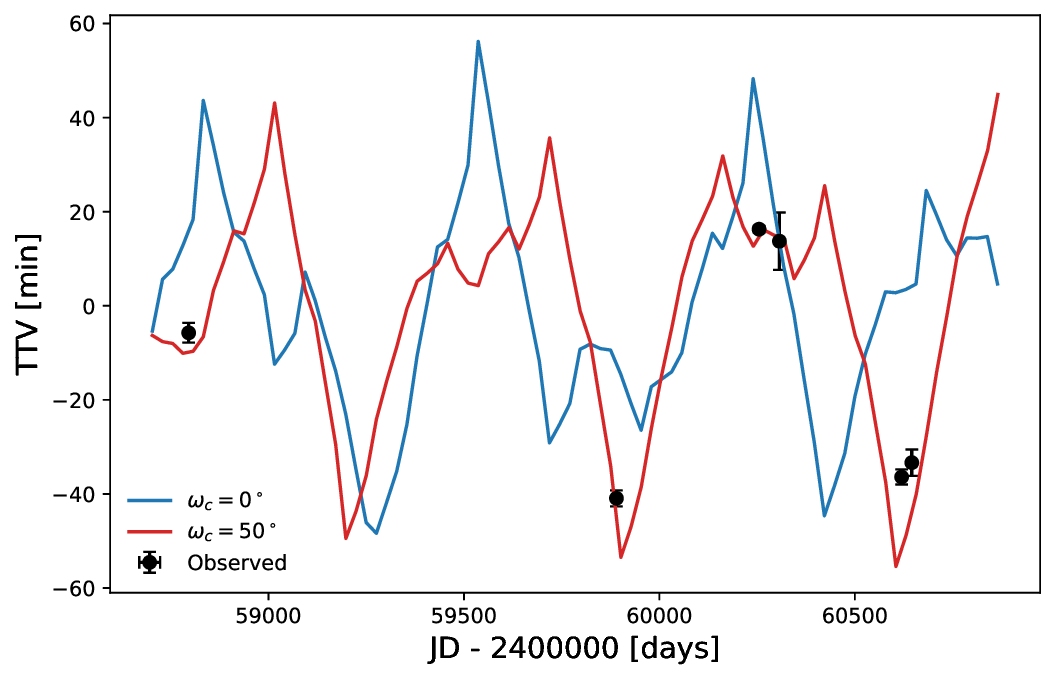}
    \caption{Transit-timing variations of the TOI-6041b planet. We adopt $e_\mathrm{c} = 0.3$ and two values of the argument of pericenter: $\omega_\mathrm{c} = 0^\circ$ (blue) and $\omega_\mathrm{c} = 50^\circ$ (red), while all other parameters are fixed at their nominal values from Table~\ref{toi6041_result_exofast}. The dots correspond to the observed TTVs. While the amplitudes are consistent with observations, the signal shape depends strongly on the orbital phase angles.}
    \label{fig:ttv_phase_sensitivity}
\end{figure}

In contrast, increasing the eccentricity of planet c had a stronger effect on the TTVs: for $e_\mathrm{c} \sim 0.3$ (i.e. the largest stable value, see Section \ref{stability}), the simulated TTV amplitudes become comparable to those observed. However, the shape of the signal becomes highly sensitive to the initial orbital phase angles, particularly the argument of pericenter $\omega_\mathrm{c}$ and mean longitude $\lambda_\mathrm{c}$. Figure~\ref{fig:ttv_phase_sensitivity} illustrates this degeneracy by showing two configurations with identical eccentricity but different $\omega_\mathrm{c}$ values. This highlights a key limitation of TTV modeling: different combinations of poorly constrained orbital elements can yield similar TTV signatures \citep{Lithwick2012}. It is worth noting that the period ratio between planets b and c is close to a 7:2 commensurability, corresponding to a fifth-order mean-motion resonance (MMR). Such a high-order resonance would typically generate weaker TTV signals unless the perturber has a relatively large eccentricity, consistent with the value required in our modeling. Moreover, in observed systems, TTVs are often associated with first- or second-order MMRs, which makes it less likely that planet c alone is responsible for the observed TTVs for planet b. Additional photometric observations will be needed to improve the analysis of the TTVs and assess the viability of this configuration.

\subsection{Effect of a Potential Third Planet}
A tentative 11 d signal in our RV-only analysis presented in Sect. \ref{rv_only}, suggests the possible presence of an additional inner planet. To evaluate its potential dynamical impact, we included a hypothetical third planet in our numerical simulations.  We assumed a circular orbit with $P = 11$ d, a RV semi-amplitude of $K = 0.5$ m/s, and all orbital angles set to $0^\circ$, except for the inclination ($i = 90^\circ$). We first verified that the system remains dynamically stable in the presence of this third planet. To investigate this, we varied its semi-major axis and eccentricity over a wide range and performed a stability analysis. The stability map (Figure.~\ref{fig:stab_40d}) reveals that stable orbits are indeed possible for semi-major axes below $0.15$ AU, corresponding to orbital periods shorter than $22$ d.\\

\begin{figure}[t!]
\centering
\includegraphics[width=\columnwidth]{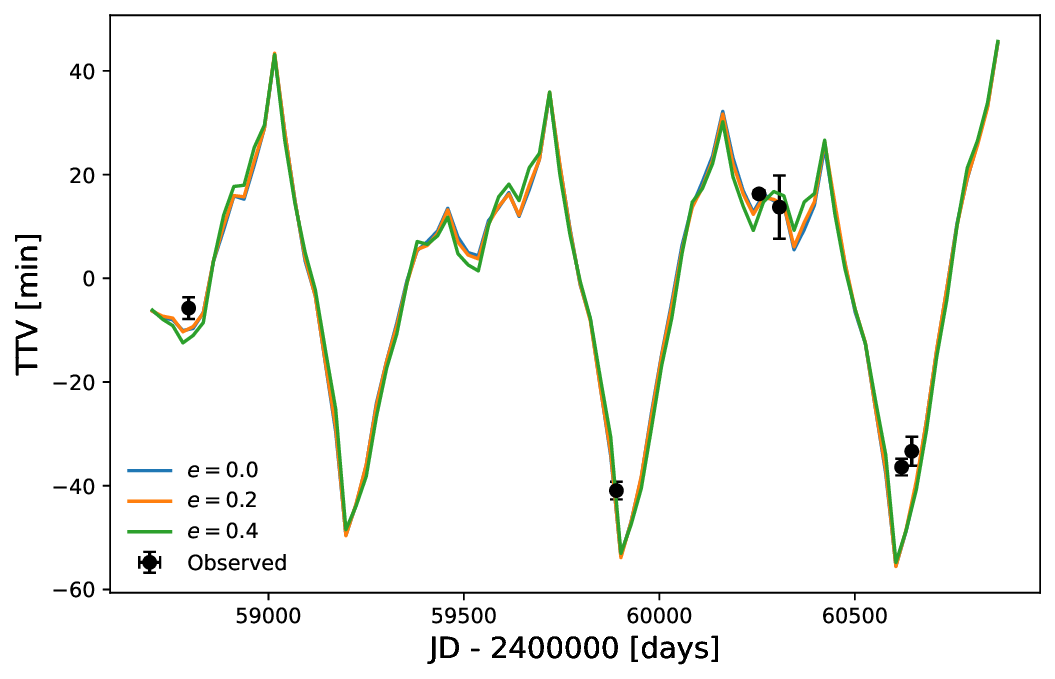}
\caption{Transit-timing variations of TOI-6041b in a three-planet model including the hypothetical 11 d planet. We adopt the same favorable configuration as in Figure~\ref{fig:ttv_phase_sensitivity}, with $e_\mathrm{c} = 0.3$ and $\omega_\mathrm{c} = 50^\circ$, and vary the eccentricity of the additional planet $e$ from $ 0$ to $0.4$. The resulting curves show negligible differences, confirming that planet c likely is the dominant source of the observed TTVs in this configuration.}
\label{fig:ttv_3rd}
\end{figure}

Next, we tested its impact on the TTVs of planet b. In the circular and coplanar configuration for all three planets, the additional planet’s influence is negligible: the resulting TTVs remain below 2 minutes. Its effect remained negligible across the different orbital configurations we explored. For example, in Figure~\ref{fig:ttv_3rd}, we adopt the favorable setup previously shown to reproduce the observed TTVs—specifically, $e_\mathrm{c} = 0.3$ and $\omega_\mathrm{c} = 50^\circ$ for planet c—and include the additional planet while varying its eccentricity $e$ between $ 0$ and $0.4$. The resulting TTV signal of planet b remains essentially unchanged.

We further tested several near-resonant periods for this additional inner planet, placing it closer to low-order MMRs with planet b (e.g. 3:2, 4:3, and 5:4). Such resonances could enhance the gravitational interactions with planet b and potentially account for the large observed TTV amplitudes. Specifically, we ran numerical simulations with the third planet placed at orbital periods of 17.3, 19.63, and 20.96 d, and computed the TTVs of planet b, assuming circular and coplanar orbits for all three planets. As shown in Figure~\ref{fig:ttv_3rdinnerplanet}, the resulting TTV amplitudes can be comparable to those observed, particularly for the periods near the 4:3 and 5:4 MMRs. In this configuration, the additional planet acts as the main perturber inducing TTVs on planet b, while planet c contributes negligibly. The exact shape of the signal remains sensitive to the poorly constrained orbital elements and the inherent degeneracies of TTV modeling. Nevertheless, this provides a plausible explanation for the observed TTVs of planet b that does not require a high eccentricity for planet c.

We also explored the possibility of a third undetected planet located between planets b and c, assuming the same RV semi-amplitude ($K = 0.5$ m/s) as before. As shown in Figure~\ref{fig:stab_40d}, stable orbits exist in this region for low eccentricities, particularly near 0.25 AU, corresponding to orbital periods between approximately 30 and 70 d. This range includes several possible MMRs: for instance, a planet with a period around 34.7 d lies near the 4:3 MMR with planet b and the 5:2 MMR with planet c, whereas a period around 39 d corresponds to the 3:2 MMR with planet b and the 9:4 MMR with planet c. 

To test its impact on the TTVs of planet b, we placed the third planet at orbital periods of 36 and 41 d and computed the TTVs of planet b, again assuming circular and coplanar orbits for all three planets. As shown in Figure~\ref{fig:ttv_3rdplanet2}, these configurations also produce TTV amplitudes of the same order as the observed variations. In this case, the additional planet acts as the main perturber inducing TTVs on planet b, though planet c also contributes non-negligibly to the signal. As before, the detailed signal shape remains sensitive to the poorly constrained orbital elements and the inherent degeneracies of TTV modeling. Nevertheless, this scenario offers another  plausible explanation for the observed variations. It does not require a high eccentricity for planet c and is consistent with the expectation that the dynamically stable region between planets b and c could host an additional, undetected planet. New transit observations and RV measurements will be essential to confirm the presence of such possible companions and to refine the TTV modeling. 

\section{Summary and discussion}
\label{discussion}

TOI-6041 is a G7-type star with a radius of $R_* = 0.879 \pm 0.026~R_{\odot}$, a mass of $M_* = 0.885^{+0.042}_{-0.036} ~M_{\odot}$, and is located at a distance of $72.3 \pm 0.1$\,pc. The system hosts at least two planets. The inner planet, TOI-6041\,b, is a transiting Neptune-like planet with a period of $26.04945^{+0.00033}_{-0.00034}$ d, a radius of $4.55^{+0.18}_{-0.17}\,R_\oplus$, and a $3\sigma$ upper limit on the RV semi-amplitude of 7\,m\,s$^{-1}$, corresponding to a maximum mass of $28.9\,M_\oplus$. This constraint implies a bulk density below approximately $1.7~\mathrm{g\,cm^{-3}}$, suggesting the presence of a substantial gaseous envelope. Further RV measurements would enhance the precision of the inner planet’s mass estimate and enable modeling of its internal structure.

The outer companion, TOI-6041\,c, is detected only through our RV measurements and has a minimum mass of $0.245^{+0.029}_{-0.028}~M_{\mathrm{Jup}}$ and an orbital period of $88.0^{+1.6}_{-1.3}$\,d. Preliminary dynamical modeling of the observed TTVs of TOI-6041\,b suggests that they could arise from gravitational perturbations by planet\,c, although matching the observed amplitudes requires a relatively high eccentricity of about 0.3. By integrating the dynamical stability analysis with the observed minimum mass constraints, we derive a 1$\sigma$ upper bound on the true mass of planet c of approximately $0.80~M_{\mathrm{Jup}}$. An alternative is the presence of a third, low-mass planet located between planets\,b and\,c, or on an inner orbit relative to planet b---particularly near a mean-motion resonance with planet\,b---which could significantly contribute to the observed variations. In this three-planet configuration, the observed TTV amplitudes can be reproduced without requiring large eccentricities.

With a magnitude of $V = 9.84 \pm 0.03$\,mag, TOI-6041 is a good candidate for follow-up observations. Using the $3\sigma$ upper mass limit of the inner planet, we estimate a lower bound on the Transmission Spectroscopy Metric \citep[TSM;][]{kempton2018framework} of TSM\,$\geq$\,59, suggesting that TOI-6041\,b is a good target for atmospheric characterization. This is particularly compelling given its estimated equilibrium temperature of $605.6^{+8.4}_{-8.3}$\,K, which places the planet near the so-called ``hazy atmosphere'' regime (270--600\,K) proposed by \citet{yu2021haze}. This offers an opportunity to probe haze formation and to characterize atmospheric compositions above it \citep{kawashima2019detectable}. Moreover, cooler atmospheres may exhibit disequilibrium chemistry, providing additional insight into atmospheric processes and their underlying physics \citep{fortney2020beyond}.

In parallel, TOI-6041\,b is well suited for spin–orbit alignment measurements, which can help constrain its dynamical history. Using Equation~6 of \citet{gaudi2007prospects}, we estimated a Rossiter--McLaughlin \citep[RM;][]{mclaughlin1924some, rossiter1924detection} semi-amplitude of $7.7 \pm 2.4$\,m\,s$^{-1}$—within reach of high-precision spectrographs such as HARPS-N. Moreover, the relatively high impact parameter of the transit (b= $0.74 \pm 0.02$) is advantageous for RM observations, as it reduces the degeneracy between $v\sin i_\star$ and the projected spin–orbit angle $\lambda$ \citep[see][]{gaudi2007prospects}, thereby allowing for a more precise determination of $\lambda$. Planets on relatively long-period orbits, such as TOI-6041\,b, are particularly valuable for RM observations, as their spin--orbit angles are less likely to have been modified by tidal realignment \citep{li2016tidal}, and therefore may retain signatures of their formation and migration histories.

\begin{figure}
\centering
\includegraphics[width=\columnwidth]{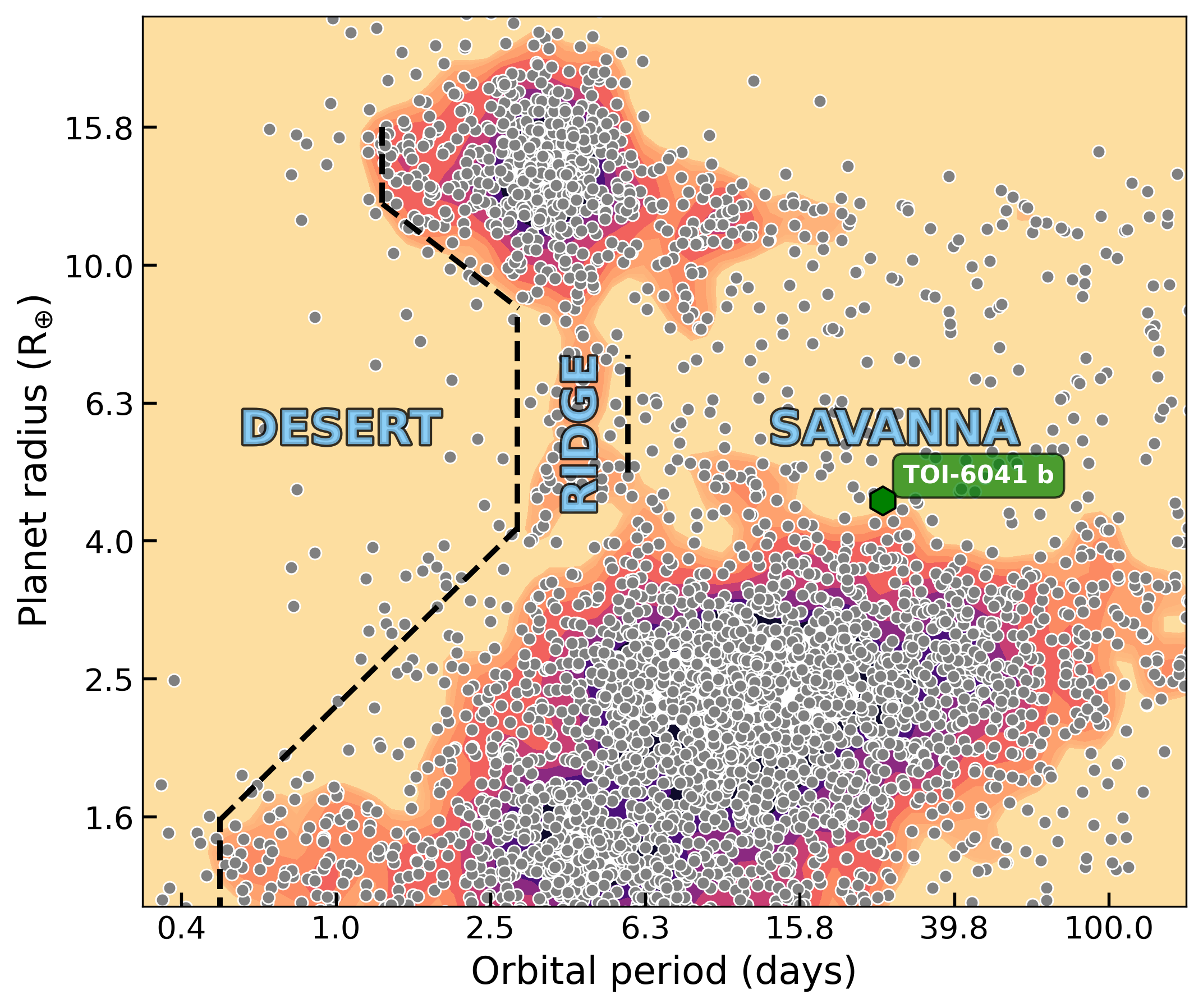}
\caption{Location of TOI-6041~b within the radius--period diagram, along with other known exoplanets whose radius precision is better than 30\%, taken from the NASA Exoplanet Archive (\url{https://exoplanetarchive.ipac.caltech.edu/}) as of June 5, 2025. The background density map illustrates the distribution of confirmed exoplanets, with regions labeled as the Desert, Ridge, and Savanna to reflect differing planet occurrence rates \citep{2024A&A...689A.250C}. TOI-6041~b (green marker) lies in the Savanna region. This plot was generated using the \texttt{nep-des} Python code (\url{https://github.com/castro-gzlz/nep-des}).}
\label{fig:savanna}
\end{figure}

The location of TOI-6041\,b in the so-called Neptunian ``savanna'' further strengthens the scientific interest of this system (see Fig. \ref{fig:savanna}). This region of parameter space corresponds to a milder deficit of intermediate-sized planets at longer orbital periods and lower levels of stellar irradiation \citep{bourrier2023dream}, in contrast to the more pronounced Neptunian desert at shorter periods ($P \lesssim 4$\,d; \citealt{mazeh2016dearth}). While the origin of these features in the radius--period plane is not fully understood, several physical mechanisms have been proposed.

Runaway gas accretion, which is predicted to occur beyond the ice line, is one such mechanism thought to contribute to the gap between sub-Neptunes and gas giants \citep{1996Icar..124...62P,2009Icar..199..338L}. The presence of TOI-6041\,b and other similar planets, located well within the ice line, therefore raises the question of whether it formed in situ or instead migrated inward after formation. If migration occurred, the nature of the process—whether through smooth disk-driven migration or high-eccentricity dynamical interactions—could leave observable signatures, such as spin--orbit misalignments or non-zero orbital eccentricities. Continued characterisation of the system, including improved mass constraints, eccentricity measurements, and RM observations, could thus provide valuable information on the migration pathways of Neptunian planets in this transitional regime.

Finally, the significant TTVs observed in this system offer a valuable opportunity to study planet--planet gravitational interactions. Additional photometric and RV observations will also be essential to refine the TTV analysis and to better characterize the architecture and interactions within the system.

\bibliography{aanda}

\begin{acknowledgements}
N.H. acknowledges support from a postdoctoral fellowship funded by the Domaine de Recherche et d’Innovation Majeur (DIM), financed by the Île-de-France Region. This work was granted access to the HPC resources of MesoPSL financed by the Region Île-de-France and the project Equip@Meso (reference ANR-10-EQPX-29-01) of the  programme Investissements d’Avenir supervised by the Agence Nationale pour la Recherche. We warmly thank the OHP staff for their support on the observations. We received funding from the French Programme National de Physique Stellaire (PNPS) and the Programme National de Planétologie (PNP) of CNRS (INSU). This paper made use of data collected by the TESS mission which is publicly available from the Mikulski Archive for Space Telescopes (MAST) operated by the Space Telescope Science Institute (STScI). Funding for the TESS mission is provided by NASA’s Science Mission Directorate. We acknowledge the use of public TESS data from pipelines at the TESS Science Office and at the TESS Science Processing Operations Center. Resources supporting this work were provided by the NASA High-End Computing (HEC) Program through the NASA Advanced Supercomputing (NAS) Division at Ames Research Center for the production of the SPOC data products. Some of the observations in this paper made use of the High-Resolution Imaging instrument ‘Alopeke and were obtained under Gemini LLP Proposal Number:
GN/S-2021A-LP-105. ‘Alopeke was funded by the NASA Exoplanet Exploration Program and
built at the NASA Ames Research Center by Steve B. Howell, Nic Scott, Elliott P.
Horch, and Emmett Quigley. Alopeke was mounted on the Gemini North telescope of the
international Gemini Observatory, a program of NSF’s OIR Lab, which is managed by
the Association of Universities for Research in Astronomy (AURA) under a cooperative
agreement with the National Science Foundation. on behalf of the Gemini partnership:
the National Science Foundation (United States), National Research Council (Canada),
Agencia Nacional de Investigación y Desarrollo (Chile), Ministerio de Ciencia,
Tecnología e Innovación (Argentina), Ministério da Ciência, Tecnologia, Inovações e
Comunicações (Brazil), and Korea Astronomy and Space Science Institute (Republic of
Korea).  the catalog described here is a subset of the full AllWISE catalog available from IRSA (http://irsa.ipac.caltech.edu/) with a selection of the columns of the full catalog. D. D. and Z. E. acknowledge support from the TESS Guest Investigator Program grant 80NSSC23K0769.

This work has been carried out within the framework of the NCCR PlanetS supported by the Swiss National Science Foundation under grants 51NF40\_182901 and 51NF40\_205606. MNG is the ESA CHEOPS Project Scientist and Mission Representative. BMM is the ESA CHEOPS Project Scientist. All of them are/were responsible for the Guest Observers (GO) Programme. None of them relay/relayed proprietary information between the GO and Guaranteed Time Observation (GTO) Programmes, nor do/did they decide on the definition and target selection of the GTO Programme.
TWi acknowledges support from the UKSA and the University of Warwick. YAl acknowledges support from the Swiss National Science Foundation (SNSF) under grant 200020\_192038. C.B. acknowledges support from the Swiss Space Office through the ESA PRODEX program.; This work has been carried out within the framework of the NCCR PlanetS supported by the Swiss National Science Foundation under grants 51NF40\_182901 and 51NF40\_205606. ACMC acknowledges support from the FCT, Portugal, through the CFisUC projects UIDB/04564/2020 and UIDP/04564/2020, with DOI identifiers 10.54499/UIDB/04564/2020 and 10.54499/UIDP/04564/2020, respectively.;A.C., A.D., B.E., K.G., and J.K. acknowledge their role as ESA\-appointed CHEOPS Science Team Members. DG gratefully acknowledges financial support from the CRT foundation under Grant No. 2018.2323 “Gaseousor rocky? Unveiling the nature of small worlds”. S.G.S. acknowledge support from FCT through FCT contract nr. CEECIND/00826/2018 and POPH/FSE (EC); The Portuguese team thanks the Portuguese Space Agency for the provision of financial support in the framework of the PRODEX Programme of the European Space Agency (ESA) under contract number 4000142255. DB, EP, EV, IR and RA acknowledge financial support from the Agencia Estatal de Investigación of the Ministerio de Ciencia e Innovación MCIN/AEI/10.13039/501100011033 and the ERDF “A way of making Europe” through projects PID2021\-125627OB\-C31, PID2021\-125627OB\-C32, PID2021\-127289NB\-I00, PID2023\-150468NB\-I00 and PID2023\-149439NB\-C41; from the Centre of Excellence “Severo Ochoa'' award to the Instituto de Astrofísica de Canarias (CEX2019\-000920\-S), the Centre of Excellence “María de Maeztu” award to the Institut de Ciències de l’Espai (CEX2020\-001058\-M), and from the Generalitat de Catalunya/CERCA programme. SCCB acknowledges the support from Fundação para a Ciência e Tecnologia (FCT) in the form of work contract through the Scientific Employment Incentive program with reference 2023.06687.CEECIND and DOI 10.54499/2023.06687.CEECIND/CP2839/CT0002. LBo, GBr, VNa, IPa, GPi, RRa, GSc, VSi, and TZi acknowledge support from CHEOPS ASI\-INAF agreement n. 2019\-29\-HH.0. ABr was supported by the SNSA. ACC acknowledges support from STFC consolidated grant number ST/V000861/1, and UKSA grant number ST/X002217/1. P.E.C. is funded by the Austrian Science Fund (FWF) Erwin Schroedinger Fellowship, program J4595\-N. This project was supported by the CNES. A.De. acknowledges financial support from the Swiss National Science Foundation (SNSF) for project 200021\_200726. This work was supported by FCT \- Fundação para a Ciência e a Tecnologia through national funds and by FEDER through COMPETE2020 through the research grants UIDB/04434/2020, UIDP/04434/2020, 2022.06962.PTDC.; O.D.S.D. is supported in the form of work contract (DL 57/2016/CP1364/CT0004) funded by national funds through FCT. B.\-O. D. acknowledges support from the Swiss State Secretariat for Education, Research and Innovation (SERI) under contract number MB22.00046. A.C., A.D., B.E., K.G., and J.K. acknowledge their role as ESA\-appointed CHEOPS Science Team Members. This project has received funding from the Swiss National Science Foundation for project 200021\_200726. It has also been carried out within the framework of the National Centre of Competence in Research PlanetS supported by the Swiss National Science Foundation under grant 51NF40\_205606. The authors acknowledge the financial support of the SNSF. MF and CMP gratefully acknowledge the support of the Swedish National Space Agency (DNR 65/19, 174/18). M.G. is an F.R.S.\-FNRS \textbf{Research Director}. CHe acknowledges financial support from the Österreichische Akademie 1158 der Wissenschaften and from the European Union H2020\-MSCA\-ITN\-2019 1159 under Grant Agreement no. 860470 (CHAMELEON). Calculations were performed using supercomputer resources provided by the Vienna Scientific Cluster (VSC). K.W.F.L. was supported by Deutsche Forschungsgemeinschaft grants RA714/14\-1 within the DFG Schwerpunkt SPP 1992, Exploring the Diversity of Extrasolar Planets. This work was granted access to the HPC resources of MesoPSL financed by the Region Ile de France and the project Equip@Meso (reference ANR\-10\-EQPX\-29\-01) of the programme Investissements d'Avenir supervised by the Agence Nationale pour la Recherche.
This work has been carried out within the framework of the NCCR PlanetS supported by the Swiss National Science Foundation under grants 51NF40\_182901 and 51NF40\_205606. AL acknowledges support of the Swiss National Science Foundation under grant number TMSGI2-211697. ML acknowledges support of the Swiss National Science Foundation under grant number PCEFP2\_194576.
PM acknowledges support from STFC research grant number ST/R000638/1. This work was also partially supported by a grant from the Simons Foundation (PI Queloz, grant number 327127).
NCSa acknowledges funding by the European Union (ERC, FIERCE, 101052347). Views and opinions expressed are however those of the author(s) only and do not necessarily reflect those of the European Union or the European Research Council. Neither the European Union nor the granting authority can be held responsible for them. A. S. acknowledges support from the Swiss Space Office through the ESA PRODEX program. V.V.G. is an F.R.S\-FNRS Research Associate.
JV acknowledges support from the Swiss National Science Foundation (SNSF) under grant PZ00P2\_208945. NAW acknowledges UKSA grant ST/R004838/1. This research made use of \texttt{nep-des} (available in \url{https://github.com/castro-gzlz/nep-des}). GyMSz acknowledges the support of the Hungarian National Research, Development and Innovation Office (NKFIH) grant K\-125015, a a PRODEX Experiment Agreement No. 4000137122, the Lend\"ulet LP2018\-7/2021 grant of the Hungarian Academy of Science and the support of the city of Szombathely. The Belgian participation to CHEOPS has been supported by the Belgian Federal Science Policy Office (BELSPO) in the framework of the PRODEX Program, and by the University of Liège through an ARC grant for Concerted Research Actions financed by the Wallonia-Brussels Federation. AL acknowledges support of the Swiss National Science Foundation under grant number TMSGI2-211697. AL and JK acknowledge support of the Swiss National Science Foundation under grant number TMSGI2-211697.

\end{acknowledgements}

\begin{appendix}

\section{TESS Target Pixel File}

\begin{figure}[h!]
\centering
\includegraphics[width=\columnwidth]{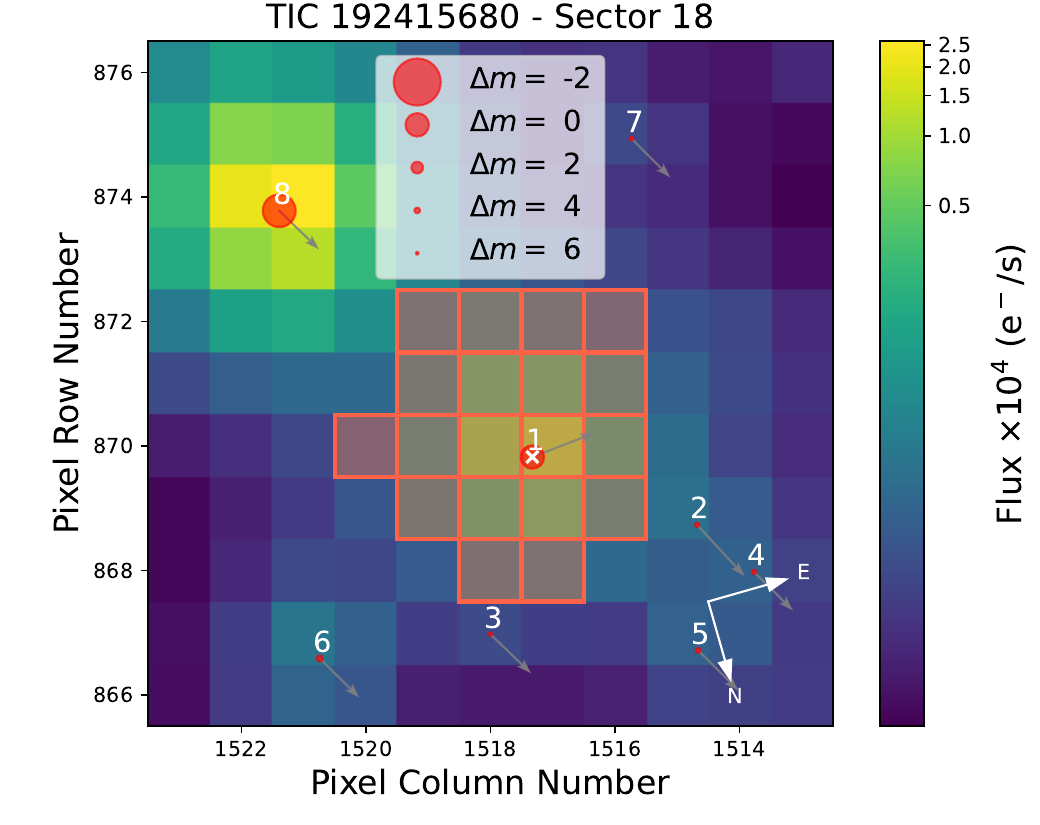}
\caption{\textit{TESS} Sector 18 target pixel file image, produced with \texttt{tpfplotter} \citep{2020AA...635A.128A}. The image display electron counts, with red-contoured pixels indicating the SPOC aperture mask. Red circles mark the primary target (labeled as 1) and nearby sources (labeled with other numbers) at their \textit{Gaia} DR3 positions. The size of each circle represents the relative brightness of the corresponding source compared to the target star. Arrows illustrate the proper motion of each star.}
\label{tpfplotter}
\end{figure}

\section{Period alias analysis}

\begin{figure}[h!]
    \centering
    \includegraphics[width=0.6\columnwidth]{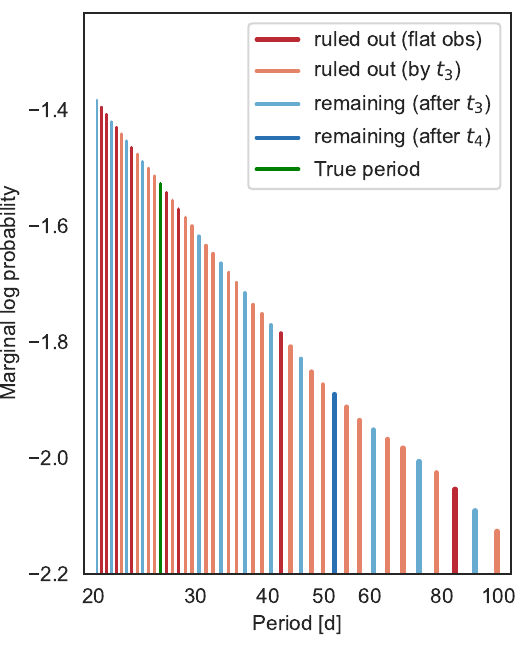}
    \caption{Log probabilities for individual aliases below 100~d from \texttt{MonoTools} analysis. Dark red bars show aliases ruled out by \textit{CHEOPS} non-detections. Light red bars show aliases ruled out by the initial \textit{CHEOPS} transit (at $t_{3}=2460254.82$; also identical to the alias constraints from the additional \textit{TESS} transit at $t_{5}=2460619.47$). Light blue bars show potential aliases after the third transit. The dark blue bar ($P_b\sim52$\,d) shows the only other remaining alias after the second \textit{CHEOPS} transit detection (at $t_{4}=2460306.92$), and the green bar shows the true period confirmed by the third \textit{CHEOPS} transit observation (at $t_{6}=2460645.53$). Only solutions with $P<100$ d are shown.}
    \label{fig:monotools}
\end{figure}

\section{High resolution image}

\begin{figure}
\centering
\includegraphics[width=0.9\columnwidth]{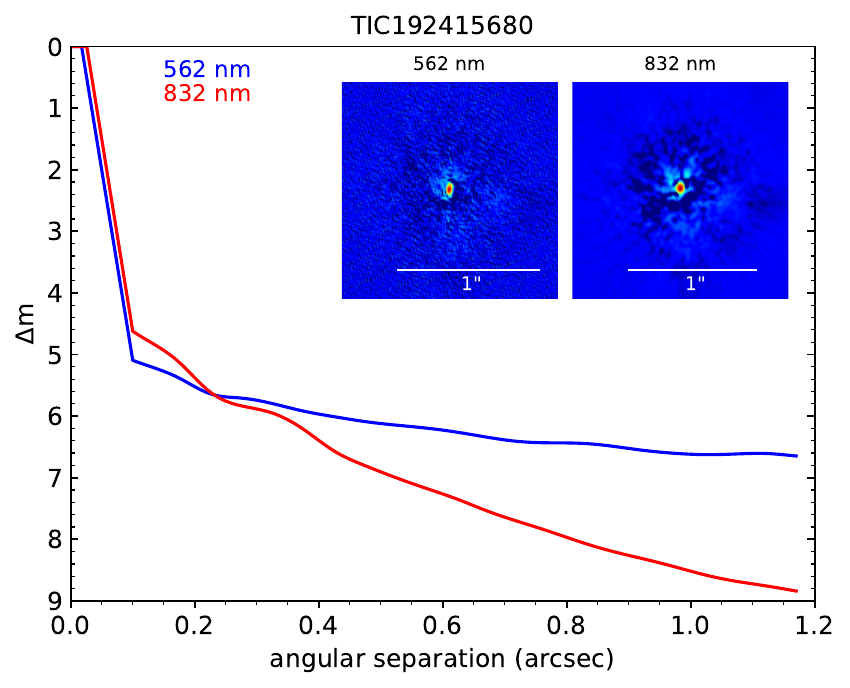}
\caption{The contrast curves for speckle interferometry observations of TOI-6041 are represented by blue and red solid lines, corresponding to the 562 nm and 832 nm bands, respectively. The plot also features the final reconstructed images, which are exhibited in the upper right corner.}
\label{image}
\end{figure}

\twocolumn
\section{RV time series}
\begin{table}[h!]
\caption{APF RVs for TOI-6041}
\label{tab:rvs APF}
\resizebox{0.46\textwidth}{!}{%
\begin{tabular}{lllll}
\hline
\hline
BJD ($-2400000$ d) & RV (m\,s$^{-1}$) & $\sigma_\mathrm{RV}$ (m\,s$^{-1}$) & S-index & $\sigma_\mathrm{S-index}$ \\

\hline
58878.746286 & -3.675772  & 3.705683 & 0.285060 & 0.002000 \\
58879.726339 & -8.463644  & 3.290296 & 0.281664 & 0.002000 \\
58879.738838 & -10.997050 & 3.874931 & 0.286212 & 0.002000 \\
58880.797159 & -9.933371  & 3.128346 & 0.332416 & 0.002000 \\
58880.812424 & -2.336793  & 3.193655 & 0.365784 & 0.002000 \\
58881.670902 & -1.405450  & 2.060878 & 0.274256 & 0.002000 \\
58884.751924 & 2.814829   & 3.092716 & 0.329735 & 0.002000 \\
58886.765389 & -4.769997  & 2.528419 & 0.315359 & 0.002000 \\
58913.656211 & 3.664502   & 2.145155 & 0.286027 & 0.002000 \\
59065.005918 & -17.797050 & 1.920321 & 0.290075 & 0.002000 \\
59151.875374 & 3.007125   & 1.906633 & 0.258785 & 0.002000 \\
59175.81492  & 10.663193  & 1.952455 & 0.301509 & 0.002000 \\
59178.799161 & 9.402857   & 2.273092 & 0.268886 & 0.002000 \\
59183.839845 & 12.460935  & 1.684850 & 0.253937 & 0.002000 \\
59185.745209 & 16.822493  & 1.802879 & 0.255540 & 0.002000 \\
59187.824279 & 21.703418  & 2.133267 & 0.299052 & 0.002000 \\
59189.721148 & 25.945226  & 2.426826 & 0.277322 & 0.002000 \\
59192.811932 & -0.260808  & 2.545385 & 0.240542 & 0.002000 \\
59194.730117 & 23.689326  & 2.835935 & 0.232222 & 0.002000 \\
59196.746152 & 17.039138  & 2.876056 & 0.235607 & 0.002000 \\
59203.745308 & 6.798748   & 2.251436 & 0.254774 & 0.002000 \\
59205.809787 & 5.332593   & 3.136430 & 0.271199 & 0.002000 \\
59220.708426 & 0.493159   & 1.988439 & 0.271507 & 0.002000 \\
59222.726767 & -16.407493 & 1.878274 & 0.287557 & 0.002000 \\
59223.775499 & -13.654458 & 3.084584 & 0.284669 & 0.002000 \\
59224.707155 & -17.611471 & 2.228663 & 0.267820 & 0.002000 \\
59225.669654 & -15.966185 & 3.911181 & 0.304180 & 0.002000 \\
59227.782816 & -19.936158 & 2.040957 & 0.249690 & 0.002000 \\
59229.694648 & 0.122848   & 2.116640 & 0.233523 & 0.002000 \\
59232.655802 & 4.722274   & 2.326371 & 0.270615 & 0.002000 \\
59252.689723 & 15.755542  & 2.859335 & 0.246251 & 0.002000 \\
59256.652165 & 4.500011   & 2.125033 & 0.279923 & 0.002000 \\
59263.767872 & 17.587455  & 2.991848 & 0.295455 & 0.002000 \\
59264.669769 & 17.425248  & 2.796155 & 0.267046 & 0.002000 \\
59268.769189 & -20.203679 & 2.702685 & 0.229752 & 0.002000 \\
59271.77326  & -6.655588  & 2.415996 & 0.236950 & 0.002000 \\
59300.643436 & 5.779297   & 2.001258 & 0.235421 & 0.002000 \\
59302.643426 & -7.140546  & 2.161766 & 0.280446 & 0.002000 \\
59389.952101 & 6.882930   & 2.666451 & 0.329347 & 0.002000 \\
59393.957888 & -14.504518 & 2.662398 & 0.232953 & 0.002000 \\
59394.972102 & -25.068372 & 4.071342 & 0.100356 & 0.002000 \\
59396.945806 & -11.793712 & 2.117844 & 0.219664 & 0.002000 \\
59400.924655 & -12.651099 & 2.256776 & 0.195205 & 0.002000 \\
59405.922931 & -16.243811 & 2.307827 & 0.272837 & 0.002000 \\
59414.985241 & -26.136852 & 2.186811 & 0.233966 & 0.002000 \\
59416.986705 & -13.068008 & 2.131959 & 0.277445 & 0.002000 \\
59417.88872  & -19.877650 & 2.693949 & 0.227865 & 0.002000 \\
59418.989513 & -17.074286 & 2.044257 & 0.282066 & 0.002000 \\
59425.861153 & -6.294075  & 2.245834 & 0.238931 & 0.002000 \\
59435.896564 & -8.876406  & 1.854445 & 0.285025 & 0.002000 \\
59438.000685 & -4.181937  & 1.831004 & 0.254261 & 0.002000 \\
59440.003684 & 10.628007  & 2.211236 & 0.256321 & 0.002000 \\
59443.940052 & 11.270397  & 2.258986 & 0.190402 & 0.002000 \\
59452.911009 & 25.396281  & 1.897756 & 0.308554 & 0.002000 \\
59457.836395 & 8.840762   & 2.029293 & 0.270793 & 0.002000 \\
59458.84345  & 7.762176   & 2.052025 & 0.278758 & 0.002000 \\
59461.955661 & 11.806341  & 2.938594 & 0.251668 & 0.002000 \\
59463.862224 & 6.488133   & 1.981545 & 0.258889 & 0.002000 \\
59469.849794 & 18.908617  & 1.885259 & 0.260064 & 0.002000 \\
59474.86075  & 13.335184  & 1.845414 & 0.309356 & 0.002000 \\
\hline
\multicolumn{5}{|r|}{{Continued on next column}} \\ \hline
\end{tabular}%
}
\end{table}

\begin{table}[h!]
\caption*{APF RVs for TOI-6041 (continued from the previous column).}
\resizebox{0.46\textwidth}{!}{%
\begin{tabular}{lllll}
\hline
BJD ($-2400000$ d) & RV (m\,s$^{-1}$) & $\sigma_\mathrm{RV}$ (m\,s$^{-1}$) & S-index & $\sigma_\mathrm{S-index}$ \\
\hline
59484.832335 & -11.059417 & 2.101504 & 0.254515 & 0.002000 \\
59487.710553 & -15.593764 & 2.452338 & 0.258694 & 0.002000 \\
59491.844004 & 8.734471   & 2.135005 & 0.323412 & 0.002000 \\
59497.825031 & -19.616376 & 2.038971 & 0.292149 & 0.002000 \\
59500.780816 & -12.891278 & 2.061234 & 0.270509 & 0.002000 \\
59502.81946  & -11.342598 & 1.935440 & 0.239413 & 0.002000 \\
59504.81671  & -13.756204 & 2.653957 & 0.249042 & 0.002000 \\
59515.830421 & 4.940400   & 2.450592 & 0.273709 & 0.002000 \\
59521.922046 & 2.841695   & 1.904567 & 0.262198 & 0.002000 \\
59529.609743 & -7.240996  & 2.357965 & 0.231821 & 0.002000 \\
59535.657523 & -1.764802  & 3.112409 & 0.304540 & 0.002000 \\
59539.783013 & 8.323448   & 2.361199 & 0.232327 & 0.002000 \\
59543.802504 & 15.999291  & 2.526530 & 0.201152 & 0.002000 \\
59547.817931 & 25.617852  & 1.907665 & 0.255249 & 0.002000 \\
59551.787616 & 9.392959   & 2.123997 & 0.262901 & 0.002000 \\
59556.780105 & 1.649591   & 3.957952 & 0.239870 & 0.002000 \\
59565.958976 & 12.277812  & 2.721480 & 0.233771 & 0.002000 \\
59585.659114 & -10.616888 & 2.485141 & 0.281554 & 0.002000 \\
59597.615832 & -28.165146 & 2.133431 & 0.203673 & 0.002000 \\
59605.813312 & -7.165401  & 1.899179 & 0.406647 & 0.002000 \\
59621.697561 & 4.152944   & 2.142703 & 0.238221 & 0.002000 \\
59638.646756 & 0.345306   & 2.484259 & 0.251508 & 0.002000 \\
59668.651078 & -23.581879 & 2.956201 & 0.228605 & 0.002000 \\
59669.651475 & -15.166291 & 2.669378 & 0.230284 & 0.002000 \\
59761.961077 & -15.912920 & 2.240801 & 0.371436 & 0.002000 \\
59784.865346 & 1.301724   & 2.756187 & 0.259621 & 0.002000 \\
59814.883885 & 6.719600   & 1.782827 & 0.232911 & 0.002000 \\
59846.815453 & -17.704571 & 2.034993 & 0.266785 & 0.002000 \\
59846.815453 & -17.704571 & 2.034993 & 0.266785 & 0.002000 \\
59895.802272 & 17.331376  & 2.011618 & 0.258387 & 0.002000 \\
59929.725474 & -5.523999  & 1.750116 & 0.226264 & 0.002000 \\
59930.658269 & 4.045986   & 1.712320 & 0.244680 & 0.002000 \\
59965.685832 & 11.280608  & 1.751465 & 0.334634 & 0.002000 \\
59969.687373 & 9.881691   & 1.849688 & 0.249993 & 0.002000 \\
59972.686895 & -1.873843  & 2.257040 & 0.313591 & 0.002000 \\
59976.716474 & 10.474115  & 1.742901 & 0.256628 & 0.002000 \\
59993.64703  & 2.025093   & 1.921243 & 0.278203 & 0.002000\\
\hline
\end{tabular}%
}
\end{table}

\begin{table}
\centering
\caption{SOPHIE RVs for TOI-6041}
\label{tab:rvs SOPHIE}
\resizebox{0.35\textwidth}{!}{%
\begin{tabular}{lll}
\hline
\hline
BJD ($-2400000$ d) & RV (m\,s$^{-1}$) & $\sigma_\mathrm{RV}$ (m\,s$^{-1}$)\\
\hline
60526.62699 &7047.5 & 2.3\\
60544.61465&7035.9 &2.2\\
60563.62546&7046.5 &2.4\\
60604.57924&7069.2 &2.2\\
\hline
\end{tabular}%
}
\end{table}

\section{Periodogram on SAP TESS data}
\onecolumn

\begin{figure}
\centering
\includegraphics[width=0.48\textwidth]{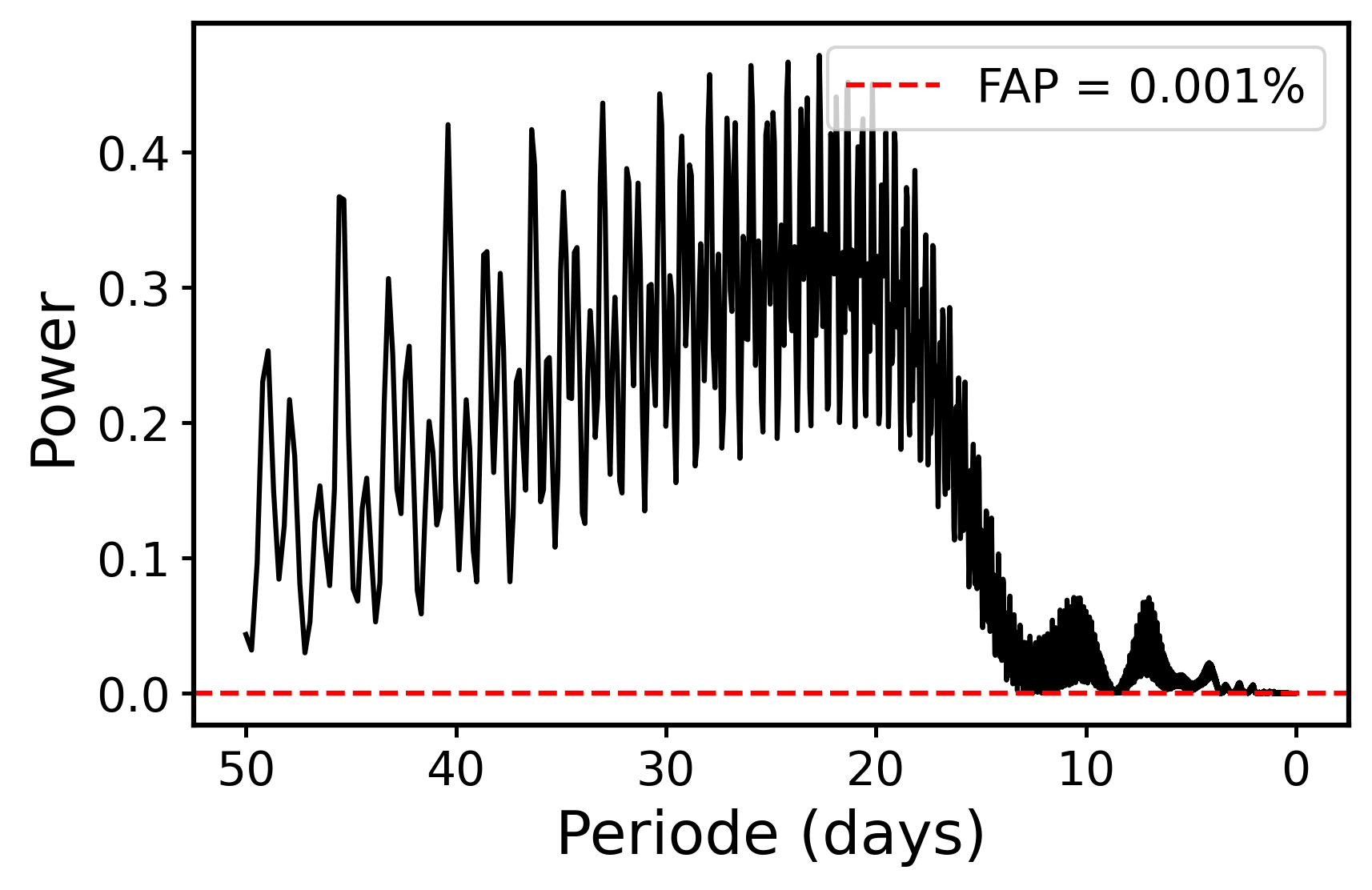}\\
\includegraphics[width=0.48\textwidth]{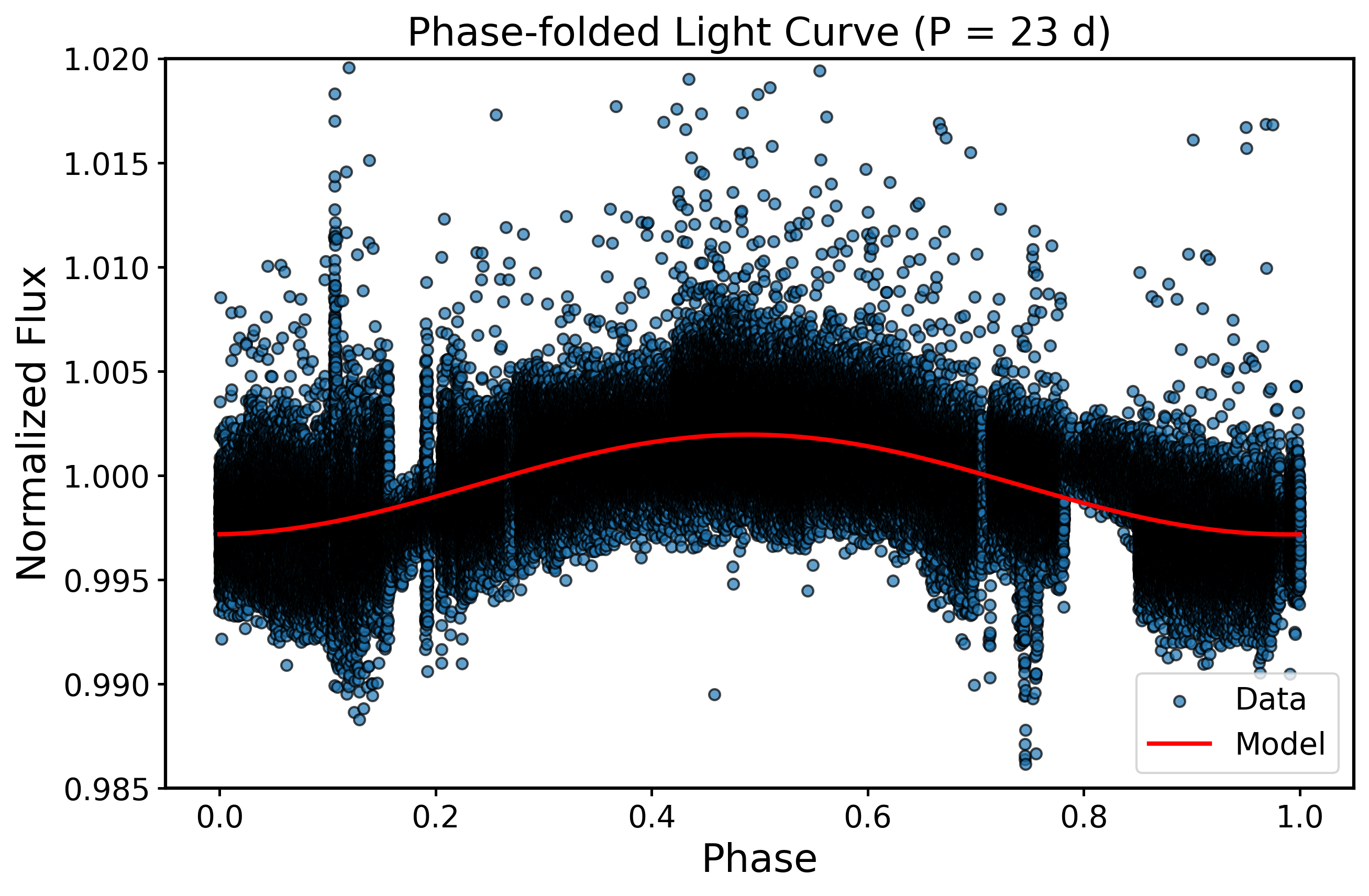}
\caption{Rotation period analysis from the SAP TESS light curve, after masking transits of TOI-6041 b. \emph{Top}: Lomb-Scargle periodogram showing a broad, significant peak between 20–30~d, under the 0.001\% FAP. \emph{Bottom}: Light curve phase-folded on the best-fit period ($\sim$23~d), with the model overplotted in red. This analysis was performed using the \href{https://docs.astropy.org}{\texttt{Astropy}} library.
}
\label{rotation_period}
\end{figure}

\section{Age probability distribution function}
\begin{figure}[h!]
\centering
\includegraphics[width=0.48\textwidth]{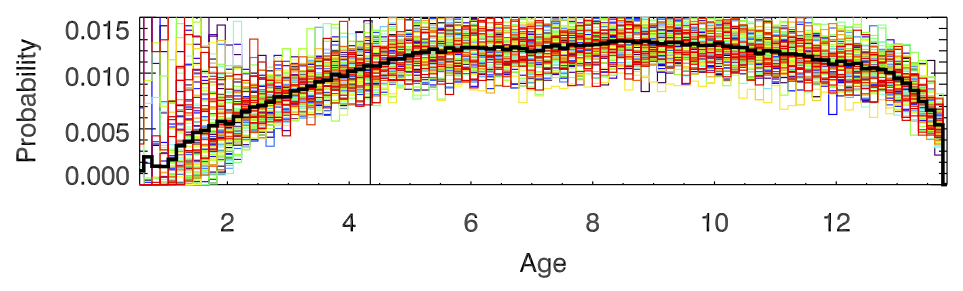}
\caption{Posterior probability distribution for stellar age from the EXOFASTv2 fit (see Section~\ref{EXOFAST}). Each color represents an individual MCMC chain, with the thick black curve showing the average across all chains. The thin black vertical line marks the best value.}
\label{age}
\end{figure}

\section{TTVs of TOI-6041 b}
\onecolumn
\begin{figure}[h!]
\centering
\includegraphics[width=0.5\columnwidth]{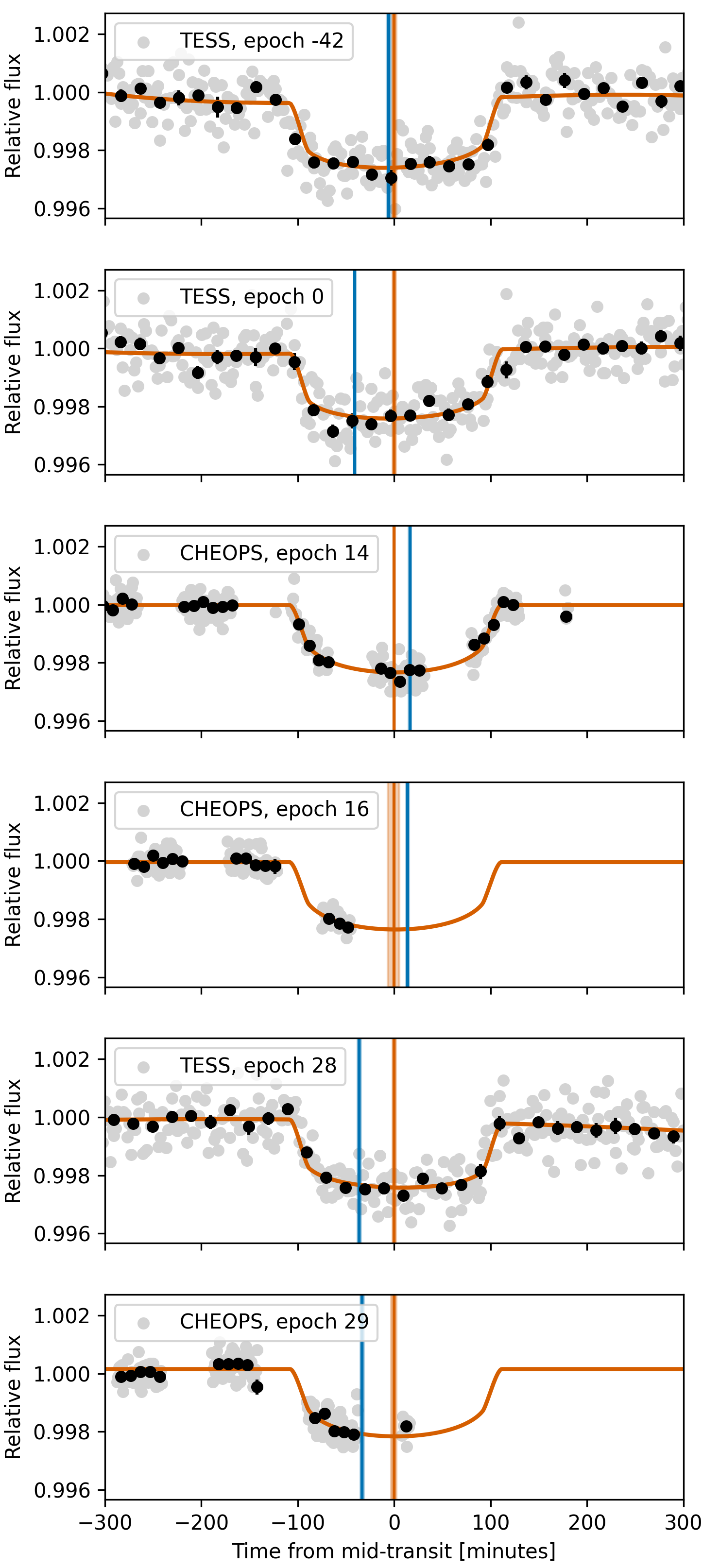}
\caption{The TOI-6041 \textit{TESS} and \textit{CHEOPS} photometric data, overplotted by the joint modeling including the TTV fit using the \texttt{EXOFASTv2}. The data points are plotted in gray, with the black points representing the binned data: 20-minute bins for the \textit{TESS} data and 9-minute bins for the \textit{CHEOPS} data. The best-fit transit models are displayed in orange. Vertical blue lines show the expected linear mid-transit times, and vertical orange lines mark the measured mid-transit times; both include their respective 1$\sigma$ error bars. The transit epoch is labeled in the top-left corner of each plot.}
\label{ttv}
\end{figure}

\begin{figure}
\centering
\includegraphics[width=0.5\columnwidth]{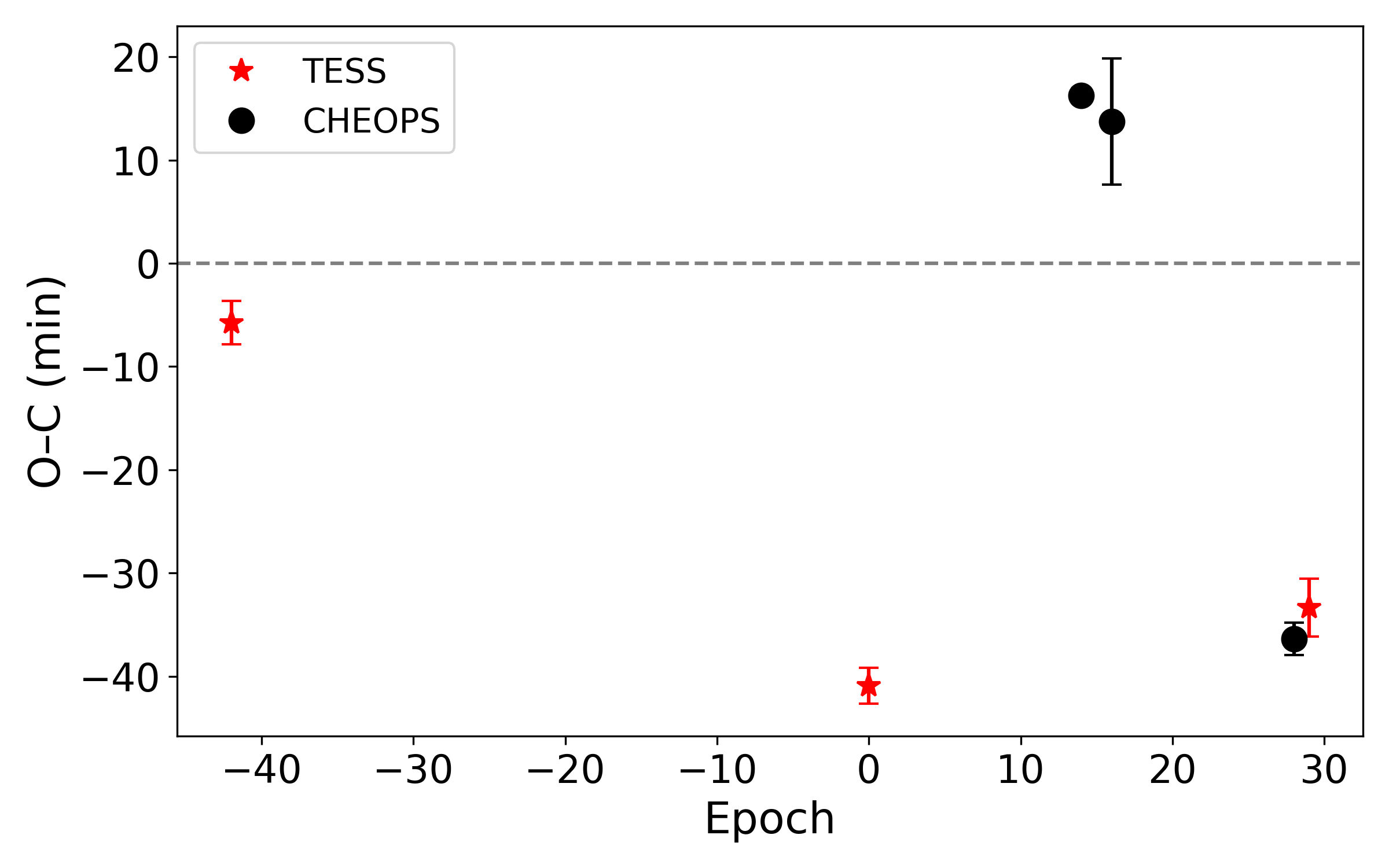}
\caption{TTV observed–calculated (O–C) diagram for TOI-6041b. The transit times are calculated using a linear ephemeris model, \( T_c = T_0 + E \cdot P \), with \( T_0 = 2459890.114417 \pm 0.000467 \) BJD and \( P = 26.049584 \pm 2.37 \times 10^{-5} \) d, where \( E \) represents the transit epoch. These values are derived by minimizing the covariance between \( T_0 \) and \( P \), following the approach described in \citet{eastman2019}.
}
\label{ttv_o_c}
\end{figure}

\section{Wavelength, telescope, and transit parameters derived from \texttt{EXOFASTv2} fit}
\begin{table*}[ht]
\centering
\resizebox{1\textwidth}{!}{%
\begin{tabular}{llccc}
\hline
\hline
Telescope Parameters: & & APF \\
\hline
$\gamma_{\rm rel}$ & Relative RV Offset (m/s) & $0.2 \pm 1.0$ \\
$\sigma_J$ & RV Jitter (m/s) & $9.51^{+0.80}_{-0.71}$\\
$\sigma_J^2$ & RV Jitter Variance & $90^{+16}_{-13}$ \\
\hline
Transit Parameters: & & & \\
\hline
\multicolumn{2}{l}{TESS Transits} 
   & Sector 18 & Sector 58 & Sector 85 \\
$\sigma^{2}$ & Added Variance 
   & $-3.4^{+3.6}_{-3.2}\times 10^{-8}$ 
   & $-0.4^{+3.4}_{-3.0}\times 10^{-8}$
   & $-5.7^{+3.1}_{-2.8}\times 10^{-8}$ \\
$F_0$ & Baseline flux 
   & $1.0002^{+0.0043}_{-0.0042}$ 
   & $1.0003 \pm 0.0042$ 
   & $1.0002 \pm 0.0043$ \\
   \\
\multicolumn{2}{l}{CHEOPS Transits} 
   & 2460254.7455 BJD & 2460306.6604 BJD & 2460645.3298 BJD \\
$\sigma^{2}$ & Added Variance 
   & $3.10^{+0.93}_{-0.83}\times 10^{-8}$
   & $3.4^{+1.2}_{-1.1}\times 10^{-8}$ 
   & $7.2^{+1.8}_{-1.6}\times 10^{-8}$ \\
$F_0$ & Baseline flux 
   & $0.999986 \pm 0.000023$ 
   & $0.999959 \pm 0.000026$
   & $1.000153^{+0.000036}_{-0.000035}$ \\
\hline
Wavelength Parameters: & & CHEOPS & TESS \\
\hline
$u_{1}$ & Linear limb-darkening coeff & $0.501 \pm 0.033$ & $0.341 \pm 0.032$ \\
$u_{2}$ & Quadratic limb-darkening coeff &$0.242^{+0.031}_{-0.032}$ & $0.233 \pm 0.030$ \\
\hline
\end{tabular}%
}
\caption{Wavelength, telescope, and transit parameters derived from \texttt{EXOFASTv2} fit.}
\label{Wavelength_exo}
\end{table*}

\section{Different tests on the effect of a potential third planet}

\begin{figure}
\centering
\includegraphics[width=0.5\columnwidth]{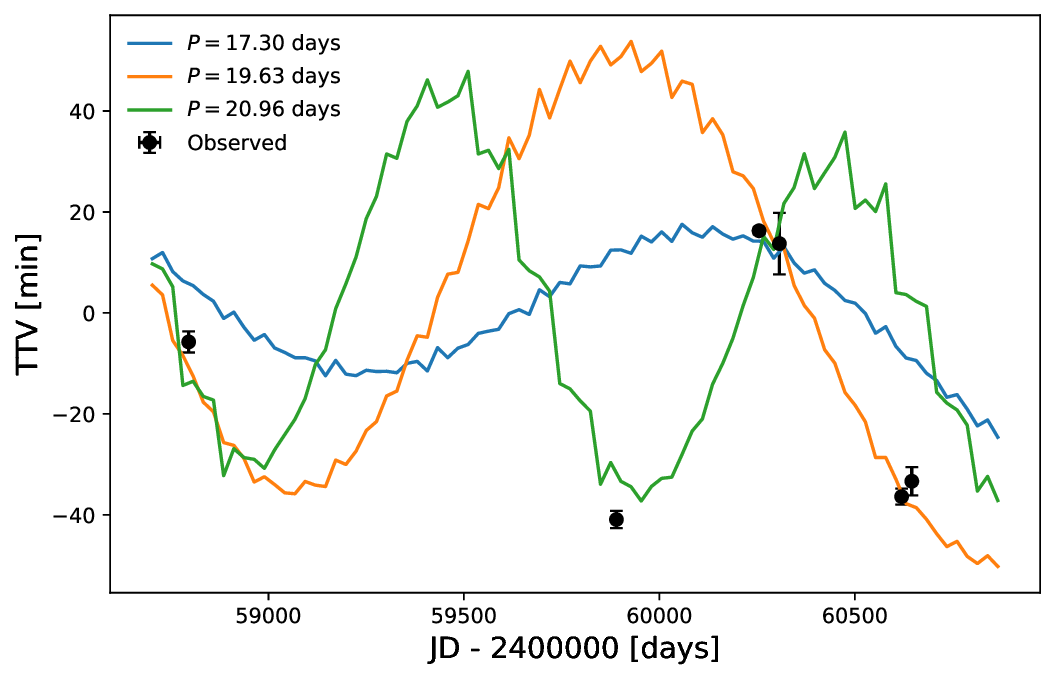}
\caption{\footnotesize{Transit-timing variations of TOI-6041b with a hypothetical planet placed interior to planet b. We adopt circular and coplanar orbits for all three planets and consider three orbital periods for the additional planet: 17.3 d (blue), 19.63 d (orange) and 20.96 d (green), corresponding to potential 3:2, 4:3 and 5:4 MMRs with planet b.}}
\label{fig:ttv_3rdinnerplanet}
\end{figure}

\begin{figure}
    \centering
    \includegraphics[width=0.5\columnwidth]{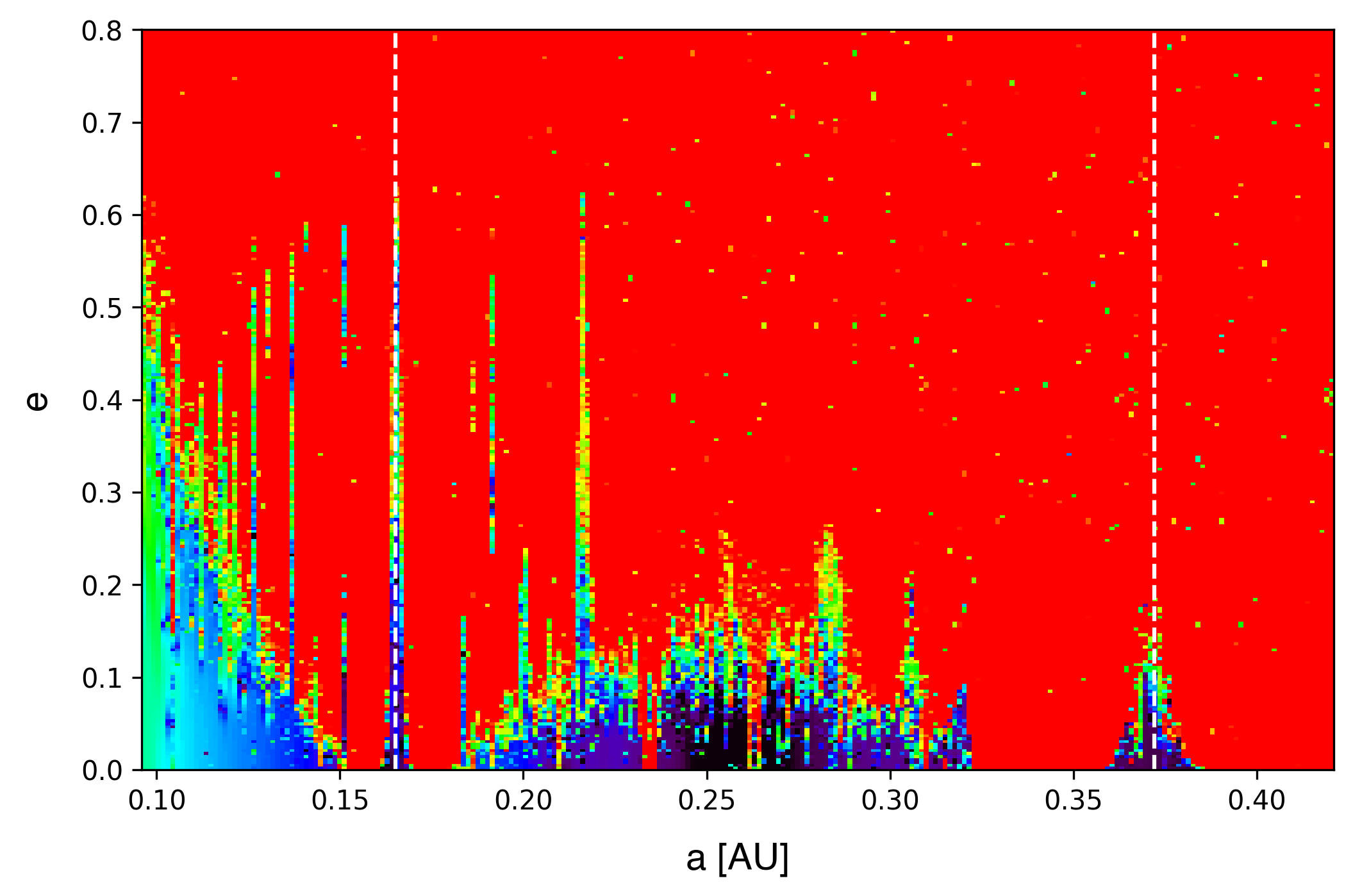}
    \caption{\footnotesize Possible location of an additional planet in the TOI-6041 system. The stability of a low-mass planet ($K = 0.5$ m/s) is analyzed as a function of semi-major axis and eccentricity. All angles are set to $0^\circ$, except for the inclination ($90^\circ$). As in Figures~\ref{fig:stability_ecc} and \ref{fig:stability_inc}, red regions indicate strongly chaotic orbits, while dark blue regions correspond to long-term stability.  White dashed lines indicate the nominal orbital periods of planets b and c.}
\label{fig:stab_40d}
\end{figure}

\begin{figure}
\centering
\includegraphics[width=0.5\columnwidth]{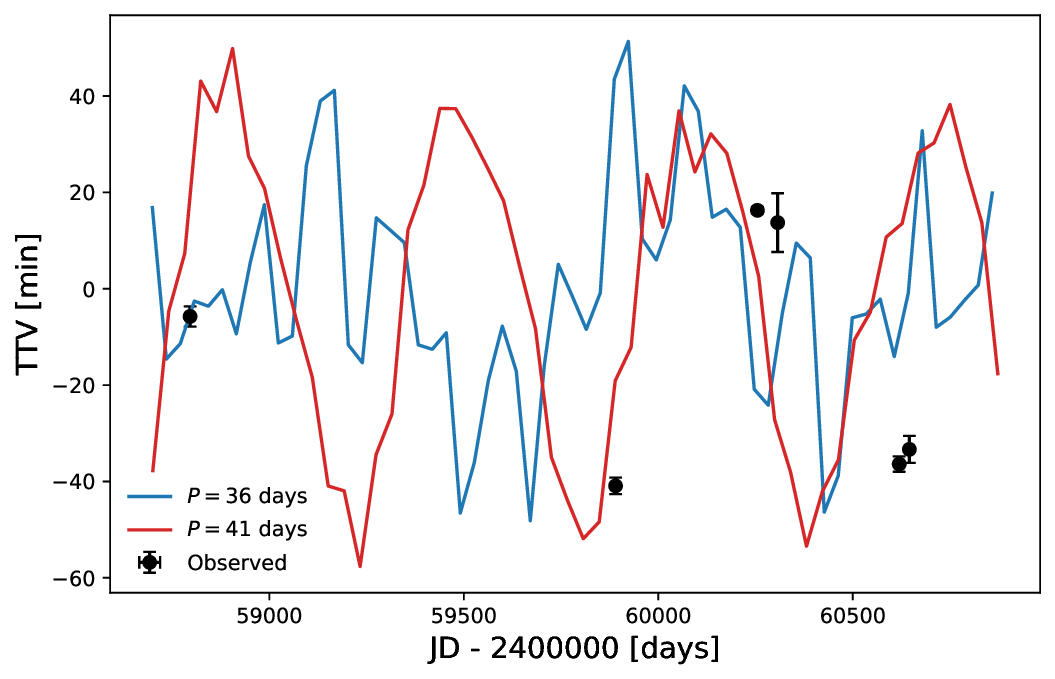}
\caption{\footnotesize{Transit-timing variations of TOI-6041b with a hypothetical planet placed between planets b and c. We adopt circular and coplanar orbits for all three planets and consider two orbital periods for the additional planet: 36 d (blue) and 41 d (red), corresponding to potential 4:3 and 3:2 MMRs with planet b. In both cases, the resulting TTV amplitudes are consistent with the observed variations.}}
\label{fig:ttv_3rdplanet2}
\end{figure}

\end{appendix}

\end{document}